%% file: main.tex
\PassOptionsToPackage{dvipsnames}{xcolor}
\documentclass[%
a4paper,%
aip,%
amsmath,amssymb,floatfix,
reprint,%
twocolumn,%
jcp,%
longbibliography,
]{revtex4-1}

\usepackage{amsmath,amsthm,latexsym,amssymb,amsfonts,epsfig}  

\usepackage{epstopdf}

\usepackage[english]{babel}
\usepackage[utf8]{inputenc}			
\usepackage{graphicx, texdraw}   %for figures
\usepackage{bm}

\usepackage{mathtools}
\usepackage{mathrsfs} 

\usepackage[outer= 1.2cm, inner = 1.3cm, top= 2cm, bottom = 1.9cm]{geometry}     % preset margins

\usepackage{enumerate}

\usepackage[colorlinks=true,citecolor=cyan, linkcolor=blue]{hyperref}

\usepackage{csquotes} 

\usepackage[usenames, dvipsnames]{color}

\usepackage{amsmath}
\usepackage{amsfonts}
\usepackage{amssymb}
\usepackage{braket}
\usepackage{lipsum}

\usepackage[normalem]{ulem} %to strike the words 
\input{commands}

\usepackage{type1ec} %

\allowdisplaybreaks

\begin{document}
\title{Hydrodynamic coupling and rotational mobilities near planar elastic membranes}

\author{Abdallah Daddi-Moussa-Ider}
\email{ider@thphy.uni-duesseldorf.de}

\affiliation{Institut f\"{u}r Theoretische Physik II: Weiche Materie, Heinrich-Heine-Universit\"{a}t D\"{u}sseldorf, Universit\"{a}tsstra\ss e 1, 40225 D\"{u}sseldorf, Germany}

\author{Maciej Lisicki}

\affiliation
{Department of Applied Mathematics and Theoretical Physics, Wilberforce Rd, Cambridge CB3 0WA, United Kingdom}

\affiliation
{Institute of Theoretical Physics, Faculty of Physics, University of Warsaw, Pasteura 5, 02-093 Warsaw, Poland }

\author{Stephan Gekle}
\affiliation
{Biofluid Simulation and Modeling, Theoretische Physik, Universit\"at Bayreuth, Universit\"{a}tsstra{\ss}e 30, 95440 Bayreuth, Germany}

\author{Andreas M. Menzel}

\affiliation{Institut f\"{u}r Theoretische Physik II: Weiche Materie, Heinrich-Heine-Universit\"{a}t D\"{u}sseldorf, Universit\"{a}tsstra\ss e 1, 40225 D\"{u}sseldorf, Germany}

\author{Hartmut Löwen}
\email{hlowen@hhu.de}

\affiliation{Institut f\"{u}r Theoretische Physik II: Weiche Materie, Heinrich-Heine-Universit\"{a}t D\"{u}sseldorf, Universit\"{a}tsstra\ss e 1, 40225 D\"{u}sseldorf, Germany}

\date{\today}

\begin{abstract}

We study theoretically and numerically the coupling and rotational hydrodynamic interactions between spherical particles near a planar elastic membrane that exhibits resistance towards shear and bending.
Using a combination of the multipole expansion and \Faxen's theorems, we express the frequency-dependent hydrodynamic mobility functions as a power series of the ratio of the particle radius to the distance from the membrane for the self mobilities, and as a power series of the ratio of the radius to the interparticle distance for the pair mobilities.
In the quasi-steady limit of zero frequency, we find that the shear- and bending-related contributions to the particle mobilities may have additive or suppressive effects depending on the membrane properties in addition to the geometric configuration of the interacting particles relative to the confining membrane.
To elucidate the effect and role of the change of sign observed in the particle self  and pair mobilities, we consider an example involving a torque-free doublet of counterrotating particles near an elastic membrane.
We find that the induced rotation rate of the doublet around its center of mass may differ in magnitude and direction depending on the membrane shear and bending properties.
Near a membrane of only energetic resistance toward shear deformation, such as that of a certain type of elastic capsules, the doublet undergoes rotation of the same sense as observed near a no-slip wall.
Near a membrane of only energetic resistance toward bending, such as that of a fluid vesicle, we find a reversed sense of rotation. 
Our analytical predictions are supplemented and compared with fully resolved boundary integral simulations where a very good agreement is obtained over the whole range of applied frequencies.

\end{abstract}
\maketitle

\section{Introduction}

The coupling between fluid flows and elastic membranes plays an important role in many physiological phenomena and is essential for understanding the biological functions and transport properties in living cells~\cite{fung13b}.
The assessment of hydrodynamic interactions between membranes and suspended tracer particles can be used as a monitor for determining the membrane mechanical properties via interfacial microrheology~\cite{gardel05, cicuta07, squires09, wirtz09}.
Such a technique has been widely employed for the measurement of the membrane viscous and elastic moduli~\cite{mason95, schnurr97, mason97, chen03}, and the characterization of the fluctuating forces in complex and active fluids~\cite{lau03, wilhelm08, foffano12}.

At small length and time scales of motion, an accurate description of the fluid flow surrounding  microscopic particles is well achieved by the linear Stokes equations~\cite{kim13}.
In these conditions, a complete description of particle motion is possible via the hydrodynamic mobility tensor, which bridges between the translational and rotational velocities of the suspended particles and the forces and torques applied on their surfaces. 
In an unbounded medium, hydrodynamic interactions are instantaneous but long-ranged, where the flow field due to a point force (Stokeslet) decays with inverse distance from the singularity position. 
However, motion in real situations often occurs in geometric confinements, where the hydrodynamic mobility is notably changed relative to the bulk value with an additional anisotropy of interactions close to boundaries\cite{happel12, dhont96}.
The need to understand and characterize these interactions has led to the development of a number of experimental techniques which allow for an accurate and reliable measurement of the particle mobility near interfaces.
Among the most popular and efficient techniques that have been utilized  are optical tweezers~\cite{meyer06,lin00, dufresne01, schaffer07}, fluorescence~\cite{kihm04, sadr05} and digital video microscopy~\cite{cui02, eral10, sharma10, dettmer14, traenkle16}, evanescent wave dynamic light scattering~\cite{michailidou09, rogers12, lisicki12}, and three-dimensional total internal reflection velocimetry techniques~\cite{huang07}.

% % Keep this paragraph please! 

% I have removed these references as maciek suggested: lorentz07, faxen22, brenner61, dean63, oneill64
% since they are cited in happel and brenner

The linearity of the Stokes equations enables the use of Green's functions to describe the flow created by an isolated point force in confined geometries, such as near a planar no-slip wall~\cite{perkins91, felderhof05, felderhof05jcp, cichocki98, swan07, swan10, franosch09}, a free interface or an interface between two immiscible fluids~\cite{lee79, lee80, berdan81, urzay07}, and also for a range of non-Cartesian geometries~\cite{fuentes88, fuentes89}. 
Analytical calculations have been carried out to include particles near interfaces with partial slip~\cite{lauga05, lauga07, felderhof12} or inside a liquid film between two fluids~\cite{felderhof06film}.
Many of the results are laid out in the monograph by Happel and Brenner~\cite{happel12}.
Additional works have examined particle dynamics near viscous interfaces~\cite{danov95b, danov98} or an interface covered with surfactant~\cite{shail83b, blawz99a, blawz10theory}.

More recently, motion of colloidal particles close to membranes with surface elasticity has attracted some attention, due to their relevance as realistic models for cell membranes~\cite{bickel06, felderhof06, bickel07, takagi11, daddi17b, daddi18epje}. Unlike fluid-solid or fluid-fluid interfaces, elastic membranes stand apart as they endow the system with memory. The motion of the particles thus depends strongly on their prior history.  
This implies the emergence of an induced long-lived subdiffusive behavior resulting from the presence of the elastic membrane in the vicinity of particles~\cite{weiss04, daddi16, daddi16b}.
Particle motion near elastic cell membranes has been experimentally investigated using optical traps~\cite{kress05, shlomovitz13, boatwright14, juenger15}, magnetic particle actuation~\cite{irmscher12}, and quasi-elastic light scattering~\cite{mizuno00, kimura05}, where a significant decrease in the mobility normal to the cell membrane has been observed in line with theoretical predictions.

In our earlier work~\cite{daddi16c}, we have studied analytically and numerically the hydrodynamic interactions between spherical particles undergoing translational motion near a planar elastic membrane.
We have found that the steady approach of two particles towards an idealized membrane with pure shear resistance may lead to attractive interactions, in contrast to the behavior known near a rigid wall where the interactions are repulsive~\cite{squires00}.

In this paper, we complete and supplement our analysis by computing the hydrodynamic coupling and rotational mobilities of a pair of particles moving near {an elastic membrane.} This is relevant to systems in which translations are restricted and the dynamics are dominated by rotational motion, such as in the case of birefringent spheres trapped in a harmonic potential interacting via their rotation-induced flow fields\cite{reichert04}.
{We thus provide the full mobility matrix for pair interactions of spheres in the presence of the elastic membrane.}
The membrane is modeled using the Skalak model~\cite{skalak73} for shear and area dilatation, and the Helfrich model~\cite{helfrich73} for bending.
We find that the contributions due to shear and bending of the particle self- and pair-mobility functions may have additive or suppressive effects depending on the membrane properties and the relative separation between the interacting particles and the membrane. 
Finally, we illustrate the physical importance of the rotational components of the self- and pair-mobility functions near a planar membrane{, as these are relevant to self-propulsion of certain types of bacterial microswimmers. In typical models of microscale swimming\cite{lauga09, lauga16}, the thrust force generated by, e.g., 	a rotating flagellum is balanced by the overall drag force on the combined cell body and the flagellum to yield the swimming speed. 
The forced rotation of the flagellum leads to a counterrotation of the cell body, together with a balancing rotational drag on the latter.	
Thus a microswimmer  is force- and torque-free. However, to evaluate the forces and torques acting on sub-elements of the swimmer, it is essential to know its mobility tensor that relates its motion to applied forces and torques. For example, rotating helicoidal flagella bundles of the bacterium \textit{E.\ coli} lead to a counterrotation of its cell body \cite{lowe87, magariyama01, macnab77, lauga06, bechinger16} to guarantee an overall torque-free nature of this suspended microswimmer in flow. It is thus particularly important to understand the coupling between the applied torque (e.g., generated by the flagellar motor) to the resulting translational velocity, which is partly motivating this work. To illustrate these ideas in practice}, we study the behavior of a torque-free doublet of two spherical particles counterrotating around their center of mass.

The remainder of the paper is organized as follows.
In Sec.~\ref{mathematicalFormulation} we present the theoretical framework we use to analytically compute the particle mobility functions by combining the multipole expansion and \Faxen's theorems for Stokes flows.
Sec.~\ref{results} provides explicit analytical expressions of the frequency-dependent coupling and rotational self and pair mobilities together with a close comparison with numerical simulations where a very good agreement is obtained.
Concluding remarks summarizing our findings and results are contained in Sec.~\ref{conclusions}. We have added Appendix~\ref{appA} containing the details of the mathematical formulation of the Green's function in the presence of a planar elastic membrane.  In Appendix~\ref{BIM} we present the completed double layer boundary integral method and the approach we have employed to numerically compute the hydrodynamic mobility functions.

\section{Mathematical model}\label{mathematicalFormulation}

\begin{figure}
\begin{center}
%\scalebox{1.}{\input{Pics/coupRotpairDiffIllustration}}
\includegraphics[scale=1]{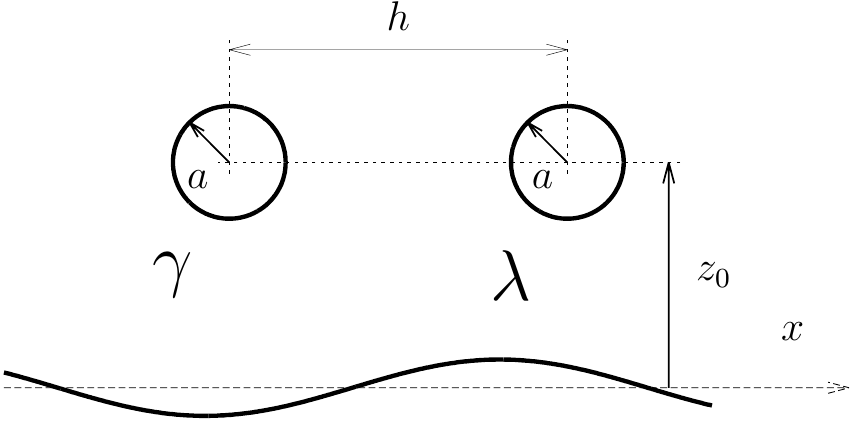}
\caption{Illustration of the system setup. A sample configuration of a pair of particles of radius~$a$, labeled as~$\gamma$ and $\lambda$, located a distance~$h$ apart from each other and a distance~$z_0$ above an elastic cell membrane. 
Here $\X_\gamma = (0,0,z_0)$ and $\X_\lambda = (h,0,z_0)$.
We define the dimensionless parameters $\epsilon := a/z_0$ and $\sigma := a/h$ characterizing the system.} 
\label{coupR_Paper_CouprototpairDiffIllustration}
\end{center}
\end{figure}

In the following, we consider two identical spherical particles of radius~$a$ immersed in a quiescent Newtonian fluid above a planar elastic membrane infinitely extended in the $xy$ plane; the $z$ direction is perpendicular to the undeformed plane.
The fluid on both sides of the membrane has the same dynamic viscosity~$\eta$, and the flow is considered incompressible.
{Both spheres feature no-slip surface conditions.}
The low-Reynolds-number hydrodynamics of a suspending incompressible fluid is governed by the forced Stokes equations~\cite{happel12}
\begin{subequations}
 \begin{align}
\eta \boldNabla^2 \bv  - \boldNabla p  + \vect{f}_1 + \vect{f}_2  &= 0 \,, \\
\boldNabla \cdot \bv  &= 0 \, , 
\end{align}
\end{subequations}
where $\bv$ and $p$ are the velocity and pressure fields, respectively.
Here $\vect{f}_\lambda$ is an arbitrary time dependent force density acting on the fluid due to the presence of particle~$\lambda$. 
{The total force and torque exerted by the spherical particle~$\lambda$ are determined by integration over its surface. Specifically,
\begin{equation}
	\vect{F}_\lambda = \oint_{S_\lambda} \vect{f}_\lambda \, \Intd S \, , \quad
	\vect{L}_\lambda = \oint_{S_\lambda} \vect{r} \times \vect{f}_\lambda \, \Intd S \, .     
\end{equation}}

If we combine the forces and torques exerted on the fluid by the particles into $\BF=(\BF_1,\BF_2)$ and $\BL=(\BL_1,\BL_2)$, and group the velocities into $\BV=(\BV_1,\BV_2)$ and $\bm{\Omega}=(\bm{\Omega}_1,\bm{\Omega}_2)$, the $12\times12$ mobility tensor  $\mi$ is defined by the relation\cite{kim13}
\begin{equation}\label{mobility_single}
   \begin{pmatrix}
   \BV \\
   \bm{\Omega} 
   \end{pmatrix} =  
   \begin{pmatrix}
   \mi^{tt} & \mi^{tr}  \\
   \mi^{rt} & \mi^{rr}  
   \end{pmatrix}
   \begin{pmatrix}
   \BF  \\
   \BL \\
   \end{pmatrix} \, .
\end{equation}
The off-diagonal components are the hydrodynamic coupling mobilities between torque and translation $(tr)$ and between force and rotation $(rt)$ and they are the transpose of each other, as required by the overall symmetry of the mobility matrix.  The mobility matrix and each of its entries can be separated into the self part that stems from the interactions of the particle with the membrane, and the pair contribution accounting in addition to the influence of the other particle. In order to determine the mobility, that is the response of the fluid to a given distribution of forces on the spheres to leading order, we now introduce the multipole expansion.

%  \subsection{Multipole expansion and \Faxen's theorem}

We consider a representative configuration of a pair {of finite-sized} particles denoted as~$\gamma$ and $\lambda$ located a distance $h$ apart from each other, and a distance $z_0$ above an elastic membrane, as schematically sketched in Fig.~\ref{coupR_Paper_CouprototpairDiffIllustration}.
In the present article, we restrict our analysis to the far-field limit, for which $a \ll z_0$.
The disturbance velocity field caused at any observation point $\vect{r}$ by a particle labeled as~$\lambda$ located at $\vect{r}_\lambda$ can be written as 
\begin{equation}
\bv (\vect{r}, \vect{r}_\lambda, \omega) = \bv^{(0)} (\vect{r}, \vect{r}_\lambda) + \bv^* (\vect{r}, \vect{r}_\lambda, \omega) \, , \label{fluidVelocitySplitUp}
\end{equation}
where $\bv^{(0)}$ denotes the fluid flow in an unbounded (infinite) fluid and $\bv^*$ is the flow field required to satisfy the boundary conditions at the membrane.

{Since the elasticity of the membrane introduces memory to the system, the response to forcing will depend on its history. As any forcing can be represented by its Fourier decomposition, we consider an oscillating force $\vect{f}_\lambda(\vect{r},\omega) = \tilde{\vect{f}}_{\lambda}(\vect{r})e^{i\omega t}$ with a characteristic frequency $\omega$. In the following, we thus work in the frequency space.} 
The disturbance field can be written as an integral over the surface of the sphere $\lambda$ as
\begin{equation}
\bv (\vect{r}, \vect{r}_{\lambda}, \omega) = \oint_{S_{\lambda}} \Gmatr (\vect{r}, \vect{r}', \omega) \cdot {\vect{f}_\lambda} (\vect{r}', \omega) \, \Intd^2 \vect{r}' \, ,
\label{fluidVelocityIntPointForce}
\end{equation}
where $\Gmatr $ denotes the velocity Green's function, i.e.\@ the flow velocity field resulting from a point force acting at position~$\vect{r}_{\lambda}$.
Similarly, the Green's function can be split up into two distinct contributions
\begin{equation}
\Gmatr (\vect{r}, \vect{r}', \omega) = \Gmatr^{(0)} (\vect{r}, \vect{r}') + \Gmatr^* (\vect{r}, \vect{r}', \omega) \, ,
\label{eqn:defDeltaG}
\end{equation}
where $\Gmatr^{(0)}$ is the infinite-space Green's function (Oseen's tensor)
\begin{equation}
\G_{\alpha \beta}^{(0)} (\vect{r},\vect{r}') = \frac{1}{8\pi\eta} \left( \frac{\delta_{\alpha \beta}}{s} + \frac{{s}_\alpha {s}_\beta}{s^3} \right) \, ,
\label{infiniteSpaceGreensFunction}
\end{equation}
with $\vect{s} :=\vect{r}-\vect{r}'$, $s:=|\vect{s}|$, and $\delta_{\alpha \beta}$ the Kronecker tensor.
The second term $\Gmatr^*$ represents the frequency-dependent correction to the Green's function due to the presence of the elastic membrane.  

Far away from the particle~$\lambda$, the integration vector variable $\vect{r}'$ in Eq.~\eqref{fluidVelocityIntPointForce} can be expanded around the particle center $\vect{r}_{\lambda}$ following a multipole expansion approach. 
{Expanding in surface moments of the force density, and truncating at the leading Stokeslet level, the disturbance velocity reads~\cite{swan07}}
\begin{equation}
\begin{split}
 \bv (\vect{r}, \vect{r}_{\lambda},\omega) &\approx \left( 1+\tfrac{a^2}{6} \boldNabla_{\vect{r}_{\lambda}}^2  \right) \Gmatr (\vect{r}, \vect{r}_{\lambda}, \omega) \cdot {\vect{F}} (\omega) \\
    &\quad\,\,+ \frac{1}{2} \,  \boldNabla_{\vect{r}_\lambda} \times \Gmatr (\vect{r}, \vect{r}_{\lambda}, \omega) \cdot {\vect{L}} (\omega) \, ,
\end{split}
\label{flowField_MultipoleExp}
\end{equation}
wherein $\boldNabla_{\vect{r}_{\lambda}}$ stands for the gradient operator taken with respect to the singularity position $\vect{r}_\lambda$, and the curl of a given tensor $\Tmatr$ is calculated as~\cite{wajnryb13}
\begin{equation}
 \left( \boldNabla \times \Tmatr \right)_{\alpha\beta} = \epsilon_{\alpha\mu\nu} \partial_{\mu} \mathcal{T}_{\nu\beta} \, ,
\end{equation}
with $\epsilon_{\alpha\mu\nu}$ being the Levi-Civita tensor. 
Note that for a single sphere in bulk, the flow field given by Eq.~\eqref{flowField_MultipoleExp} satisfies exactly the no-slip boundary conditions at the surface of the sphere~\cite{kim06}.
Using \Faxen's theorems~\cite{durlofsky87}, the translational and rotational velocities of the particle~$\gamma$ in this flow reads~\cite{swan07, swan10}
\begin{subequations}\label{Faxen_Both}
 \begin{align}
\vect{V}_{\gamma} (\omega) &= \mu_0^{tt} {\vect{F}_{\gamma}} (\omega) + \left( 1+\tfrac{a^2}{6} \boldNabla_{\vect{r}_{\gamma}}^2 \right) \bv (\vect{r}_{\gamma}, \vect{r}_{\lambda}, \omega)  \, , \label{Faxen_trans} \\
\vect{\Omega}_{\gamma} (\omega) &= \mu_0^{rr} {\vect{L}_{\gamma}} (\omega) + 
\frac{1}{2} \,  \boldNabla_{\vect{r}_\gamma} \times \bv (\vect{r}_{\gamma}, \vect{r}_{\lambda}, \omega) \, ,
\label{Faxen_rot}
\end{align}
\end{subequations}
where $\mu_0^{tt} := 1/(6\pi\eta a)$ and $\mu_0^{rr} := 1/(8\pi\eta a^3)$ denote the translational and rotational bulk mobilities, respectively.
We further emphasize that the disturbance flow $\bv$ incorporates both the disturbance from the particle~$\lambda$ in addition to that caused by the presence of the membrane.
By inserting Eq.~\eqref{flowField_MultipoleExp} into \Faxen's formulas stated by Eqs.~\eqref{Faxen_Both}, the frequency-dependent translational, coupling, and rotational pair-mobility tensors  can be calculated as
\begin{subequations}
 \begin{align}
    \mi^{tt, \gamma\lambda} (\omega) &= \left( 1+\tfrac{a^2}{6} \boldNabla_{\vect{r}_{\gamma}}^2 \right) 
    \left( 1+\tfrac{a^2}{6} \boldNabla_{\vect{r}_{\lambda}}^2  \right) \Gmatr (\vect{r}_{\gamma}, \vect{r}_{\lambda}, \omega)  \, , \label{particlePairMobility_tran}   \\ 
    \mi^{tr, \gamma\lambda} (\omega) &= \tfrac{1}{2} \, \left( 1+\tfrac{a^2}{6} \boldNabla_{\vect{r}_{\gamma}}^2 \right) {\boldNabla_{\vect{r}_\lambda}} \times \Gmatr (\vect{r}_\gamma, \vect{r}_{\lambda}, \omega) \, , \label{particlePairMobility_coupling} \\
    \mi^{rr, \gamma\lambda} (\omega) &= \tfrac{1}{4} \,  {\boldNabla_{\vect{r}_\gamma}} \times {\boldNabla_{\vect{r}_\lambda}} \times \Gmatr (\vect{r}_\gamma, \vect{r}_{\lambda}, \omega) \, . \label{particlePairMobility_rot}		
\end{align}
\end{subequations}
For the self mobilities, the correction in the flow field~$\bv^*$ 
%in Eq.~\eqref{fluidVelocitySplitUp} 
due to the presence of the second particle should be discarded as only the correction due to the presence of the membrane should be considered in \Faxen's formulas. 
Accordingly, the frequency-dependent self-mobility tensors read
\begin{subequations}
 \begin{align}
    \mi^{tt, \gamma\gamma} (\omega) &= \mu_0^{tt} \, \mathbf{1}  + \lim_{\vect{r} \to \vect{r}_\gamma} \left( 1+\tfrac{a^2}{6} \boldNabla_{\vect{r}}^2 \right)  \notag \\
    &\times  \left( 1+\tfrac{a^2}{6} \boldNabla_{\vect{r}_\gamma}^2  \right) \Gmatr^* (\vect{r}, \vect{r}_\gamma, \omega)  \, , \label{particleSelfMobility_tran}   \\ 
    \mi^{tr, \gamma\gamma} (\omega) &= \tfrac{1}{2} \, \lim_{\vect{r} \to \vect{r}_\gamma} \left( 1+\tfrac{a^2}{6} \boldNabla_{\vect{r}_{\gamma}}^2 \right) 
    {\boldNabla_{\vect{r}}} \times \Gmatr^* (\vect{r}, \vect{r}_{\gamma}, \omega) \, ,  \label{particleSelfMobility_coupling}  \\
    \mi^{rr, \gamma\gamma} (\omega) &= \mu_0^{rr} \,  \mathbf{1}  + \tfrac{1}{4} \, \lim_{\vect{r} \to \vect{r}_\gamma} 
    {\boldNabla_{\vect{r}_\gamma}} \times  {\boldNabla_{\vect{r}}} \times \Gmatr^* (\vect{r}, \vect{r}_{\gamma}, \omega) \, , \label{particleSelfMobility_rot}		
\end{align}
\end{subequations}
where $\mathbf{1}$ denotes the unit tensor. Having constructed the self- and pair-mobility tensors, the Green's functions associated with the elastic membrane need to be introduced at this point.

% \subsection{Green's function for a membrane-bounded fluid}

The exact Green's functions for a point-force acting near a planar elastic membrane has been determined in our earlier works, see e.g.\@ Refs.~\onlinecite{daddi16} and~\onlinecite{daddi17}. For completeness, we have repeated the key expressions in the Appendix~\ref{appA}. The membrane is modeled as a two dimensional sheet made of a hyperelastic material that exhibits resistance towards shear and bending.
Membrane shear elasticity is described by the Skalak model~\cite{skalak73} which is often used as a practical model for red-blood-cell membranes~\cite{foessel11, dupont15, barthes16, lac04, lim06}.  The model is characterized by the shear modulus $\kS$ and the area dilatation modulus $\kA$, which are related to each other by the coefficient $C:=\kA/\kS$.  
The strain energy for the Skalak model is given by~\cite{krueger12, kruger17}
\begin{equation}
E_\mathrm{S} = \frac{\kS}{12} \int_S \left(  I_1^2 + 2I_1 - 2I_2 +C I_2^2 \right) \Intd S \, ,
\label{W_SK}
\end{equation}
where $I_1$ and $I_2$ are the invariants of the right Cauchy-Green deformation tensor, %$C_{\alpha\beta} = F_{\gamma\alpha} F_{\gamma\beta}$, 
% where $F$ is transformation gradient tensor, 
employed in finite strain theory by Green and Adkins,~\cite{green60, zhu14,zhu15}
\begin{equation}
I_1 = G^{\alpha\beta}  g_{\alpha\beta} - 2 \, , \qquad
I_2 = \det G^{\alpha\beta}  \det g_{\alpha\beta} - 1 \, ,
\end{equation}
for $\alpha,\beta \in \{1,2\}$.
These are related to the principal inplane stretch ratios via the relations $I_1=\lambda_1^2+\lambda_2^2-2$ and $I_2 = \lambda_1^2 \lambda_2^2-1$.
Here $g_{\alpha\beta}$ are the covariant components of the metric tensor in the deformed state, and $G^{\alpha\beta}$ are the corresponding contravariant components in the undeformed state.
{Using the more familiar Lam\'{e} coefficients for a homogeneous and isotropic material in the small-strain regime, it follows that $\kS = \frac{3}{2} h \mu$ and $\kA = h \lambda$, where $h$ denotes the membrane thickness~\cite{skalak73}.}

The resistance towards bending is modeled by the Helfrich model~\cite{helfrich73,seifert97}, with the corresponding bending modulus~$\kB$. 
Accordingly, the bending energy is described by a quadratic curvature-elastic continuum model of the form~\cite{Guckenberger_preprint}
\begin{equation}
	E_\mathrm{B} = \int_S 2\kB \left( H-H_0 \right)^2 \, \Intd S  \, ,  \label{HelfrichHamiltonian}
\end{equation} 
wherein $H$ denotes the mean curvature, and $H_0$ is the spontaneous curvature which is taken consistently as a planar undeformed membrane.

In this approach, the linearized traction jumps across the membrane are related at $z=0$ to its displacement field $\boldsymbol{u}$ and the dilatation $e:=\partial_x u_{x}+\partial_y u_{y}$ via \citep{daddi16}  
  \begin{align}
  [\sigma_{z\alpha}] &= -\frac{\kS}{3} \left( \Delta_\parallel u_\alpha + (1+2C) \partial_{\alpha} e \right) \,, \quad \alpha \in \{ x,y \} \, , \label{tangentialCondition}\\
  [\sigma_{zz}] &= \kB \Delta_\parallel^2 u_z \, , \label{normalCondition}
  \end{align}
where  $[f]$ denotes the jump of a given function $f$ across the membrane. Here $\Delta_\parallel := \partial^2_{x} + \partial^2_{y}$ is the two-dimensional Laplace operator along the membrane.
The components of the stress tensor of the fluid are $\sigma_{z\alpha} = -p \delta_{z\alpha} + \eta (\partial_\alpha v_{z} + \partial_z v_{\alpha})$ for  $\alpha\in \{x,y,z\}$ \citep{kim13}.

The membrane displacement field $\boldsymbol{u}$ and the fluid velocity $\vect{v}$ at the membrane are coupled by the no-slip boundary condition prescribed to leading order in deformation at the undisplaced membrane, given in the frequency space by 
\begin{equation}
{v}_\alpha = i\omega {u}_\alpha |_{z=0} \, , \quad \alpha \in \{ x,y,z \} \, ,
\end{equation}  
where $i$ is the imaginary unit such that $i^2=-1$.

%%%%%%%%%%%%%%%%%%%%%%%%%%%%%%%%%%%%%%%%%%%%%%%%%%%%%%%%%%%%%%%

\section{Results}\label{results}

In our previous work~\cite{daddi16c}, we have provided analytical expressions for the translational mobility functions for the motion near an elastic membrane. We have shown that the frequency-dependent corrections to the particle self- and pair-mobility functions can be written as a linear superposition of the contributions stemming from shear and bending resistances. 
We now complete this result by computing the leading-order translation--rotation coupling and rotational elements of the mobility matrix, both for the self and pair mobilities.

\subsection{Self mobilities}

For an isolated particle, there is no coupling between translation and rotation.
In the two-particle system, however, this coupling occurs only when considering higher-order reflections, and it is not captured in the Rotne-Prager approximation~\cite{rotne69, ermak78,zuk14}.

Mathematical expressions for the hydrodynamic coupling and rotational self-mobility corrections are expressed in terms of power series of the ratio of particle radius to the particle-membrane distance $\epsilon := a/z_0$.
We have shown that for the translational mobility corrections, the leading-order term scales as $\epsilon$. 
We will now show that the coupling and rotational self-mobility corrections scale to leading order as $\epsilon^2$ and $\epsilon^3$, respectively.

% \subsubsection{Translation-rotation coupling}

The translation-rotation coupling mobility is readily obtained after inserting the Green's functions stated in the Appendix~\ref{appA}
%defined by Eq.~\eqref{greenFunctions} 
into Eq.~\eqref{particleSelfMobility_coupling}. In the following, we scale the coupling mobilities by $\mu_0^{tr} \equiv \mu_0^{rt} = 1/(6\pi\eta a^2)$.

After computation, we find that the self-related contributions to the mobility tensor due to membrane shear and bending can explicitly be expressed as
\begin{subequations}\label{DeltaMu_TR_XY_Self}
\hspace*{-4mm}\vbox{ % a useful tool to shift equations
 \begin{align}
\frac{\mu_{xy, \mathrm{S}}^{tr, \mathrm{S}}}{\mu_0^{tr}} &= 
\frac{3 \epsilon^2}{64} \left( \beta^2(2+i\beta)\Gamma_1+i\beta-\beta^2-2 + \frac{4i\beta}{B} + \frac{3\beta^2}{B^2} \, \Gamma_2 \right)   \notag \\
&+ \left( -\frac{3}{64} + \frac{\beta}{128} \left( 2i+\beta-i\beta^2-\beta^3 \Gamma_1 \right) \right) \epsilon^4 \, , \label{DeltaMu_TR_XY_Self_She} \\
\frac{\mu_{xy, \mathrm{B}}^{tr, \mathrm{S}}}{\mu_0^{tr}} &= \left(\frac{3}{32} - \frac{i\betaB^3}{64} \left( \psi+\phi_+ \right) \right) \epsilon^2  \notag \\
&+ \left( -\frac{3}{64} + \frac{\betaB^3}{384} \left( 3i-\betaB\psi-\psi' \right) \right)\epsilon^4  \, ,  \label{DeltaMu_TR_XY_Self_Ben}
\end{align}}
\end{subequations}
where the subscripts S and B respectively stand for shear and bending, and S appearing as a superscript stands for self.
The total coupling mobility is obtained by linear superposition.
It follows from the symmetry of the mobility tensor that  $ \mu_{yx}^{tr} = -\mu_{xy}^{tr}$ and that $ \mu_{yx}^{tr} = \mu_{xy}^{rt}$.
Here $\beta := 6Bz_0 \eta \omega/\kS$ is a dimensionless frequency associated with shear resistance, where $B=2/(1+C)$, and $\betaB:=2z_0 (4\eta\omega/\kB)^{1/3}$ is a dimensionless number associated with bending~\cite{daddi16}.
Furthermore, we define the auxiliary functions 
\begin{align}
\phi_{\pm} &:=  e^{-i \overline{z_{\mathrm{B}}}} \E_1 \left(-i\overline{z_{\mathrm{B}}} \right) \pm  e^{-i z_{\mathrm{B}}} \E_1 \left(-iz_{\mathrm{B}} \right) \, , \notag \\
\psi &:= e^{-i\betaB}\E_1 (-i\betaB) \, , \notag
\end{align}
where $z_{\mathrm{B}} := j\beta_\mathrm{B}$ and $j:=e^{2i\pi/3}$ is the principal cubic-root of unity (such that $j^3=1$). 
The function $\E_n$ denotes the generalized exponential integral defined as~\cite{abramowitz72}
\begin{equation}
	\E_n(x) := \int_1^\infty t^{-n} e^{-xt} \Intd t \, .
\end{equation}
The bar designates a complex conjugate. 
We further define  
\begin{equation}
   \Gamma_1 = e^{i\beta} \E_1(i\beta) \, ,  \qquad \Gamma_2 = e^{\frac{2i\beta}{B}} \E_1 \left(\frac{2i\beta}{B}\right) \, , 
\end{equation}
and
\begin{equation}
   \psi' = \zBbar e^{-i\zBbar} \E_1 (-i\zBbar) + \zB e^{-i\zB} \E_1 (-i\zB) \, .
\end{equation}
% We note that $B=2/(1+C)$, a parameter associated with the Skalak model.
By taking the vanishing-frequency limit in Eqs.~\eqref{DeltaMu_TR_XY_Self}, the shear- and bending-related corrections for the $xy$ component of the coupling mobility read
\begin{subequations}
 \begin{align}
\lim_{\beta\to 0} \frac{\Delta\mu_{xy, \mathrm{S}}^{tr, \mathrm{S}}}{\mu_0^{tr}} &= -\frac{3}{32}\, \epsilon^2 - \frac{3}{64} \, \epsilon^4 \, , \\
\lim_{\betaB\to 0} \frac{\Delta\mu_{xy, \mathrm{B}}^{tr, \mathrm{S}}}{\mu_0^{tr}} &= \frac{3}{32} \, \epsilon^2 - \frac{3}{64} \, \epsilon^4 \, , 
\end{align}
\end{subequations}
leading to the hard-wall limit obtained upon summing up both contributions term by term, namely~\cite{swan07}
\begin{equation}
	\lim_{\beta,\betaB \to 0} \frac{ \mu_{xy}^{tr, \mathrm{S}}}{\mu_0^{tr}} = -\frac{3}{32} \, \epsilon^4 \, , \label{DeltaMu_TR_XY_Self_hard}
\end{equation}
as first computed by Goldman~\cite{goldman67a}. 
Interestingly, the leading-order terms with~$\epsilon^2$ drop out in the steady limit where the resulting correction to the coupling pair mobility scales rather as~$\epsilon^4$. Notably, the shear and bending related parts have opposite contributions to the total mobility to leading order.
This feature will play a prime role in the rotational dynamics of a torque-free doublet of particles above an elastic membrane.

\begin{figure}
\begin{center}
%\scalebox{1.05}{\input{Pics/deltaMu_TR_Self}}
\includegraphics[scale=1.05]{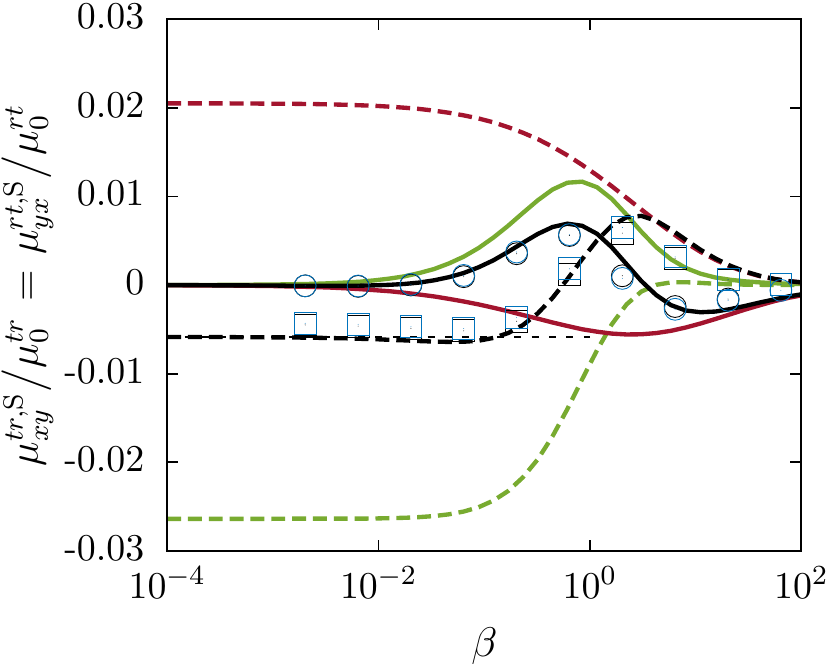}
\caption{(Color online) The scaled frequency-dependent coupling self mobility versus the scaled frequency. 
The solid particle is set a distance $z_0 = 2a$ above a planar elastic membrane, the reduced bending modulus of which is $\EB :=\kB/(\kS z_0^2)= 2/3$.
Here we take $C=1$ in the Skalak model. The theoretical predictions are shown as dashed lines for the real (reactive) part, and as solid lines for the imaginary (dissipative) part.
Symbols refer to boundary integral simulations results, where squares and circles denote the real and imaginary parts, respectively.
Overall theoretical results are shown by the black lines.
The shear/area dilatation and bending-related parts as stated by Eqs.~\eqref{DeltaMu_TR_XY_Self_She} and \eqref{DeltaMu_TR_XY_Self_Ben} are shown in green (light gray in a black-and-white printout) and red (dark gray in a black-and-white printout), respectively.
The thin horizontal dashed line stands for the coupling self mobility near a no-slip wall given by Eq.~\eqref{DeltaMu_TR_XY_Self_hard}. 
Blue symbols refer to the $rt$ component of the total mobility as obtained numerically. } 
\label{deltaMu_TR_Self}
\end{center}
\end{figure}

In Fig.~\ref{deltaMu_TR_Self}, we show the scaled coupling self mobility versus the scaled frequency $\beta$ of a particle located a distance $z_0 = 2a$ above a planar elastic membrane.
Here we consider a reduced bending modulus $\EB:=\kB/(\kS z_0^2) = 2/3$ for which the characteristic time scale for shear $T_\mathrm{S} :=6z_0\eta/\kS$ and for bending $T_\mathrm{B} := 4\eta z_0^3/ \kB$ are equal~\cite{daddi16c}.
We observe that the real and imaginary parts are nonmonotonic functions of frequency that vanish for larger frequencies, thus recovering the behavior in a bulk fluid.
In the low-frequency regime, the coupling mobility approaches that predicted near a hard wall as given by Eq.~\eqref{DeltaMu_TR_XY_Self_hard}.
Moreover, we remark that shear manifests itself in a more pronounced way than bending.
The coupling mobilities $tr$ and $rt$ as obtained numerically clearly satisfy the symmetry property required for particles in Stokes flows. 
A good agreement is obtained between theoretical predictions and boundary integral simulations over the whole range of applied frequencies.
Technical details regarding the numerical method and the procedure we have employed for the computation of the particle hydrodynamic mobilities are provided in the Appendix~\ref{BIM}.

{
It is worth mentioning that the oscillation frequency~$\omega$ of the particle should be chosen small enough for the linear response theory to be valid.
Therefore, it is essential to ensure that the Strouhal number $\operatorname{St} = a\omega/V$ satisfies $\operatorname{St} \ll 1$, where $V:=|\vect{V}|$ is the velocity amplitude. 
In typical physiological situations, $\kS \sim 10^{-6}$~N/m, $B\sim1$, and $\eta\sim 10^{-3}$~Pa$\cdot$s.
By considering a particle of radius $a=z_0/2$ with a linear velocity of $V\sim 10^{-6}$~m/s,  it follows that $\operatorname{St} \sim 10^{-4} \beta$.
In the present work, we consider a maximum scaled frequency $\beta = 10^2$ such that the condition $\operatorname{St} \ll 1$ remains always satisfied.}

\subsubsection{Rotational mobilities}

\begin{figure}
\begin{center}
%\scalebox{0.95}{\input{Pics/deltaMu_RR_Self}}
\includegraphics[scale=0.95]{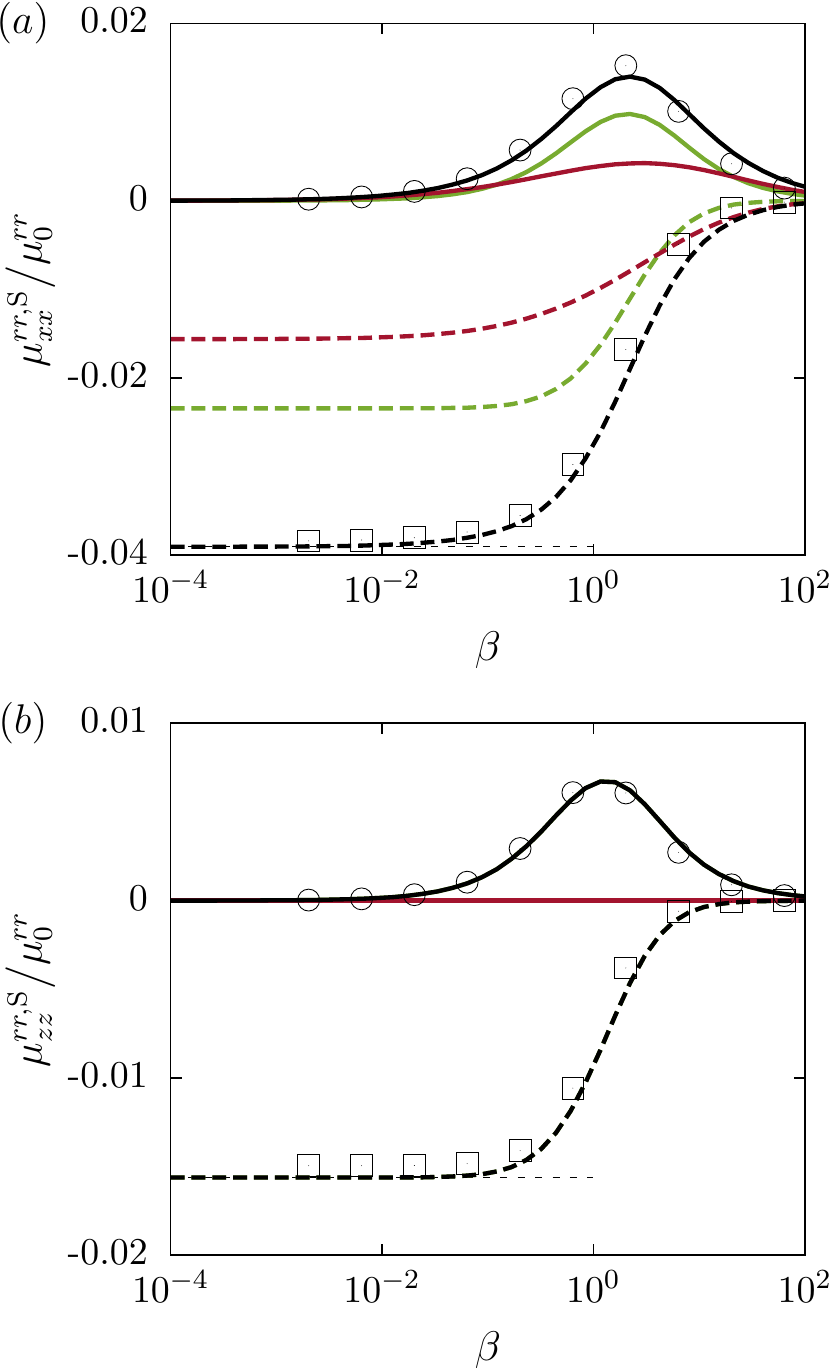}
\caption{(Color online) The scaled frequency-dependent rotational self mobility versus the scaled frequency.
The analytical predictions are given to leading order by Eqs.~\eqref{DeltaMu_RR_XX_Self} and \eqref{DeltaMu_RR_ZZ_Self}.
Here we use the same color code as in Fig.~\ref{deltaMu_TR_Self}. 
Thin horizontal dashed lines are the hard-wall predictions given by Eqs.~\eqref{DeltaMu_RR_XX_Self_hard} and \eqref{DeltaMu_RR_ZZ_Self_hard} for the components $xx$ and $zz$, respectively.} 
\label{deltaMu_RR_Self}
\end{center}
\end{figure}

The correction to the rotational mobility for the rotation around an axis parallel to the membrane is again readily obtained by inserting the Green's functions 
contained in the Appendix~\ref{appA}
%as defined in Eqs.~\eqref{greenFunctions} 
into Eq.~\eqref{particleSelfMobility_rot} to obtain
\begin{subequations}\label{DeltaMu_RR_XX_Self}
 \begin{align}
\frac{\Delta\mu_{xx, \mathrm{S}}^{rr, \mathrm{S}}}{\mu_0^{rr}} &= 
-\frac{1}{16} \bigg( i\beta^3 \left(\Gamma_1+\frac{4\Gamma_2}{B^3} \right)-\beta^2 \left( 1+\frac{2}{B^2}\right)  \notag \\
&-i\beta \left( 1+\frac{1}{B}  \right) + 3 \bigg) \epsilon^3 \, , \label{DeltaMu_RR_XX_Self_She} \\
\frac{\Delta\mu_{xx, \mathrm{B}}^{rr, \mathrm{S}}}{\mu_0^{rr}} &= 
\frac{1}{48} \left( i\betaB^3 \left(  \psi+\phi_+ \right) - 6 \right) \epsilon^3  \label{DeltaMu_RR_XX_Self_Ben}
\end{align}
\end{subequations}
for the shear and bending related parts, respectively.
Similarly, the total mobility is obtained by superposition of the contributions due to shear and bending.
The component $yy$ has an analogous expression due to the system symmetry along the horizontal plane.
In addition, for the rotation around an axis perpendicular to the membrane, the shear and bending related corrections read
\begin{subequations}\label{DeltaMu_RR_ZZ_Self}
 \begin{align}
\frac{\Delta\mu_{zz, \mathrm{S}}^{rr, \mathrm{S}}}{\mu_0^{rr}} &= 
-\frac{3iB}{16\beta} \left( 4e^{\frac{2i\beta}{B}}  \E_5 \left( \frac{2i\beta}{B} \right) - 1 \right) \epsilon^3 \, , \label{DeltaMu_RR_ZZ_Self_She} \\
\frac{\Delta\mu_{zz, \mathrm{B}}^{rr, \mathrm{S}}}{\mu_0^{rr}} &=  0 \, . \label{DeltaMu_RR_ZZ_Self_Ben}
\end{align}
\end{subequations}
Thus the rotational self mobilities have a leading-order term scaling as $\epsilon^3$. 
Furthermore, the $zz$ component depends only on the membrane shear properties and does not depend on bending.
Not surprisingly, the torque exerted on the particle along an axis normal to the planar membrane induces only an in-plane displacement of the membrane.
Therefore the resulting stresses do not cause any out-of-plane deformation or bending.
By taking the vanishing-frequency limit in the $xx$ component of the rotational mobilities in Eqs.~\eqref{DeltaMu_RR_XX_Self} we obtain
\begin{subequations}
 \begin{align}
  \lim_{\beta\to 0} \frac{\Delta\mu_{xx, \mathrm{S}}^{rr, \mathrm{S}}}{\mu_0^{rr}} &= -\frac{3}{16} \, \epsilon^3 \, , \\
  \lim_{\betaB \to 0} \frac{\Delta\mu_{xx, \mathrm{B}}^{rr, \mathrm{S}}}{\mu_0^{rr}} &= -\frac{1}{8} \, \epsilon^3 \, , 
\end{align}
\end{subequations}
leading after summing up both contributions term by term to the result near a hard wall~\cite{swan07}
\begin{equation}
\lim_{\beta,\betaB \to 0} \frac{\Delta\mu_{xx}^{rr, \mathrm{S}}}{\mu_0^{rr}} = -\frac{5}{16} \, \epsilon^3 \, . \label{DeltaMu_RR_XX_Self_hard} \\
\end{equation}
For the $zz$ component we obtain 
\begin{equation}
\lim_{\beta,\betaB \to 0} \frac{\Delta\mu_{zz}^{rr, \mathrm{S}}}{\mu_0^{rr}} =  \lim_{\beta\to 0} \frac{\Delta\mu_{zz, \mathrm{S}}^{rr, \mathrm{S}}}{\mu_0^{rr}} = -\frac{1}{8} \, \epsilon^3 \, . \label{DeltaMu_RR_ZZ_Self_hard}
\end{equation}
In the steady limit, the correction to the $xx$ component of the rotational self mobility is found to be 2.5 times larger than that of the $zz$ component.
Therefore, it is easier to rotate the particle along an axis perpendicular than parallel to a membrane endowed with a finite shear rigidity.

In Fig.~\ref{deltaMu_RR_Self}, we show the scaled rotational self mobilities versus the scaled frequency $\beta$ for the rotation about an axis parallel $(a)$ and perpendicular $(b)$ to the planar elastic membrane.
We observe that the real part is a monotonically increasing function of frequency while the imaginary part exhibits the typical peak structure which occurs at $\beta \sim 1$, which has already been seen in previous studies involving a planar membrane, particularly for the translational motion\cite{daddi16c}.

Considering the $xx$ component, we see that shear and bending both have negative  contributions to the total mobility, in contrast to the behavior observed for the coupling mobilities. The $zz$ component is solely determined by the shear resistance of the membrane while bending does not play a role for this component.
Again, the simulation results agree well with the theoretical predictions.

%%%%%%%%%%%%%%%%%%%%%%%%%%%%%%

\subsection{Pair mobilities}

\begin{figure*}
\begin{center}
% \scalebox{0.8}{\input{Pics/deltaMu_TR_Pair}}
\includegraphics[scale=0.8]{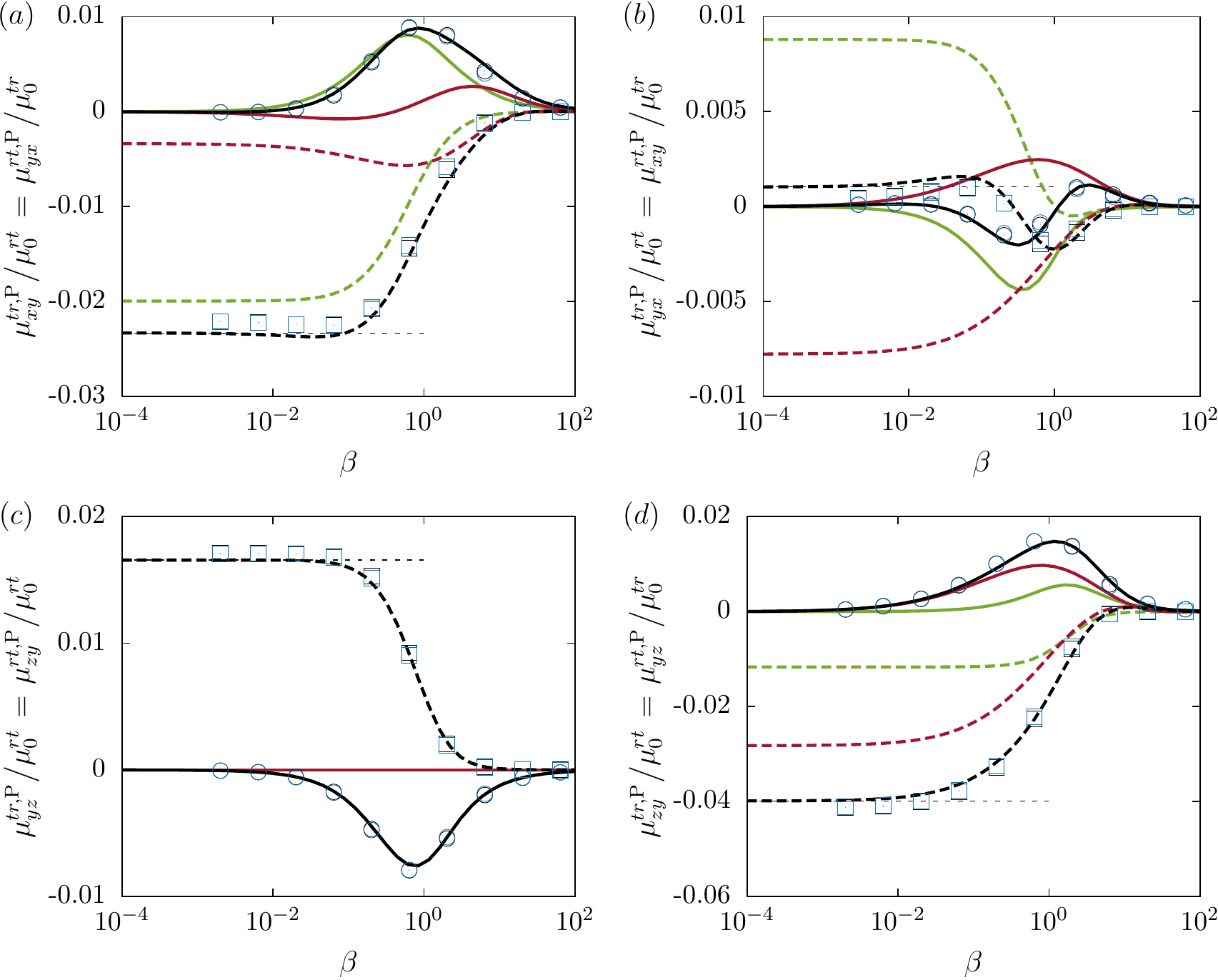}
\caption{(Color online) The scaled frequency-dependent coupling pair mobilities versus the scaled frequency as predicted theoretically by Eqs.~\eqref{DeltaMu_TR_Pair}. The color code is the same as in Fig.~\ref{deltaMu_TR_Self}. 
Here the pair is located at $z_0=2a$ above the membrane with an interparticle distance $h=4a$.
Thin horizontal dashed lines are the hard-wall predictions given by Eqs.~\eqref{DeltaMu_TR_Pair_hard}.} 
\label{deltaMu_TR_Pair}
\end{center}
\end{figure*}

\begin{figure*}
\begin{center}
% \scalebox{0.8}{\input{Pics/deltaMu_RR_Pair}}
\includegraphics[scale=0.8]{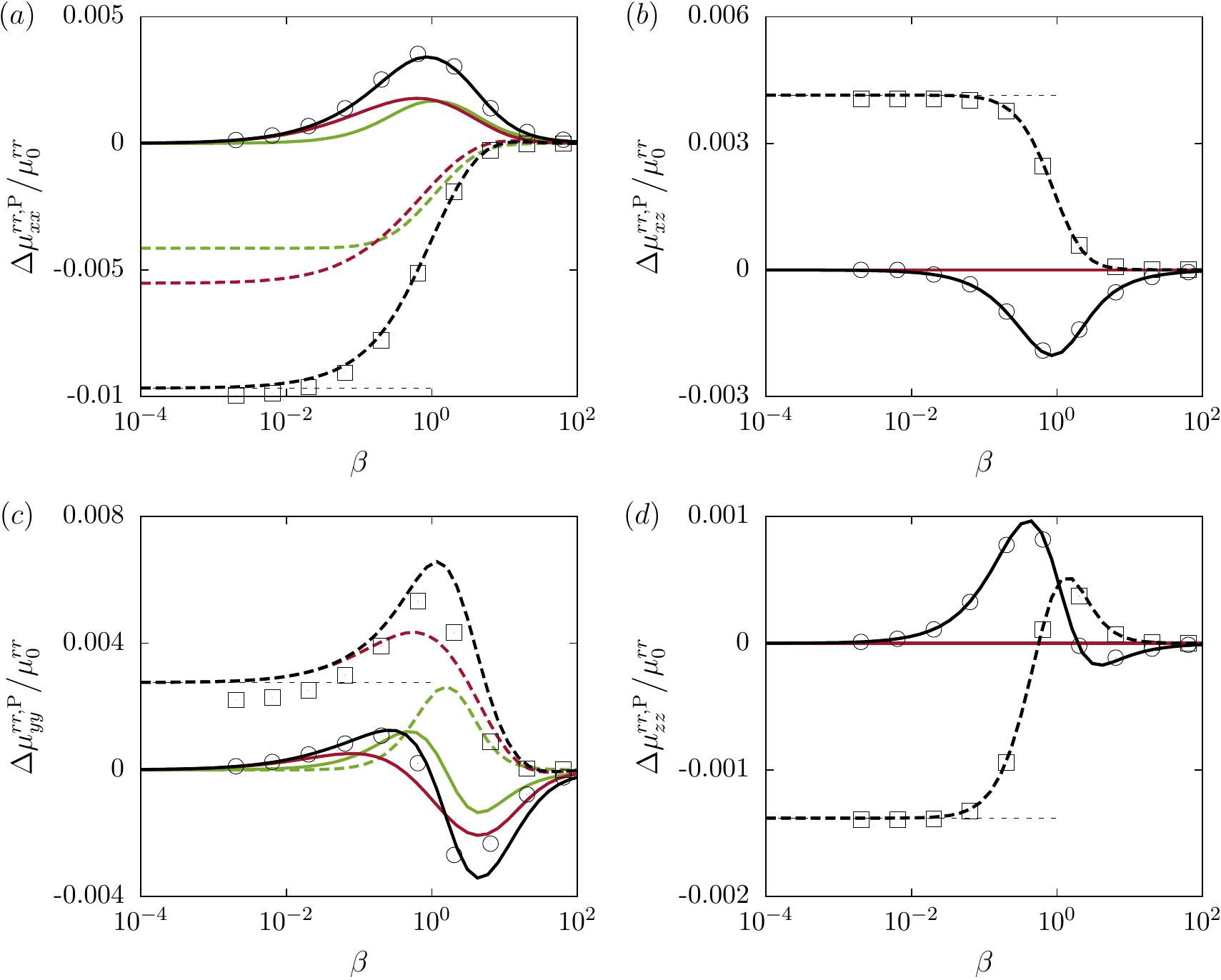}
\caption{(Color online) The scaled frequency-dependent rotational pair mobilities versus the scaled frequency as predicted theoretically by Eqs.~\eqref{DeltaMu_RR_Pair}. 
Thin horizontal dashed lines are the corrections predicted near a hard wall given by Eqs.~\eqref{DeltaMu_RR_Pair_hard}.
The color code is the same as in Fig.~\ref{deltaMu_TR_Self}.  } 
\label{deltaMu_RR_Pair}
\end{center}
\end{figure*}

Having calculated the coupling and rotational self mobilities, we now consider the fluid-mediated hydrodynamic interactions between two particles.

We express the pair mobility corrections in terms of a power series in $\sigma=a/h$. The latter takes only physical values strictly between 0 and $1/2$ to avoid overlap between the two particles.
For the translational mobility, we have shown that the leading-order corrections scale linearly with~$\sigma$. 
Similarly to self contributions, the leading-order correction terms for the coupling and rotational pair mobilities scale as $\sigma^2$ and $\sigma^3$, respectively.

%  \subsubsection{Translation-rotation coupling}

We first consider the translation--rotation coupling components of the pair-mobility tensor near an elastic membrane. 
By inserting the expressions for the Green's functions as stated  %Eqs.~\eqref{greenFunctions} 
in the Appendix~\ref{appA} into Eq.~\eqref{particlePairMobility_coupling}, the  coupling pair mobilities can be expressed in terms of convergent infinite integrals as
\begin{subequations}\label{DeltaMu_TR_Pair} 
 \begin{align}
\frac{\mu_{xy}^{tr,  \mathrm{P}}}{\mu_0^{tr}} &= \int_0^\infty  \frac{i\sigma^2 u}{\xi^{1/2}}
\Bigg(
\frac{1}{\xi^{3/2}} \bigg( \frac{\xi^{1/2}\chi_1\Gamma_+ -2u\left( 3\xi+u\Lambda \right)\chi_0}{2iu-\beta}  \notag \\
&+ \frac{4u^3 \Lambda \varphi}{8iu^3-\betaB^3} \bigg)
+\frac{3iB}{2} \frac{\chi_1}{Bu+i\beta}
\Bigg) e^{-2u} \, \Intd u \, , \label{DeltaMu_TR_XY_Pair} \\
\frac{\mu_{yx}^{tr,  \mathrm{P}}}{\mu_0^{tr}} &= \int_0^\infty \frac{\sigma^2 u}{\xi} 
\bigg( 
-\frac{3B}{2}\frac{\varphi}{Bu+i\beta} \notag \\
&+\frac{i}{\xi^{1/2}} \left( \frac{\chi_1\Gamma_+}{2iu-\beta} + \frac{4u^3 \chi_1 \Lambda}{8iu^3-\betaB^3} \right)
\bigg) e^{-2u} \, \Intd u \, , \label{DeltaMu_TR_YX_Pair} \\
\frac{\mu_{yz}^{tr, \mathrm{P}}}{\mu_0^{tr}} &= \int_0^\infty 
\frac{3B \sigma^2 u^2 \chi_1 }{\xi} \frac{e^{-2u}}{Bu+i\beta} \, \Intd u \, , \label{DeltaMu_TR_YZ_Pair} \\
\frac{\mu_{zy}^{tr, \mathrm{P}}}{\mu_0^{tr}} &= \int_0^\infty 
\frac{2i\sigma^2 u^3 \chi_1}{\xi^2} 
\left( \frac{\Lambda}{2iu-\beta} +\frac{4u \Gamma_{-}}{8iu^3-\betaB^3} \right)
e^{-2u} \Intd u \, , \label{DeltaMu_TR_ZY_Pair} 
\end{align}
\end{subequations}
where P appearing as a superscript stands for pair, and is a shorthand for the component $\gamma\lambda$.
Furthermore, we define the geometric parameter $\xi := 4z_0^2/h^2 =  4\sigma^2/\epsilon^2$ and
\begin{eqnarray}
\Lambda &:=& 4\sigma^2 u - 3\xi \, ,\notag\\
\Gamma_{\pm} &:=& 4\sigma^2 u^2-3u\xi \pm 3\xi \,,\notag\\
\chi_n &:=& J_n \left( \frac{2u}{\xi^{1/2}} \right) \, , \notag \\
\varphi &:=& \xi^{1/2}\chi_1-2u\chi_0 \, . \notag 
\end{eqnarray}
The terms involving $\beta$ and $\betaB$ in Eqs.~\eqref{DeltaMu_TR_Pair} are the contributions stemming from shear/area dilatation and bending, respectively. The component $yz$ (and thus $zy$ of the $rt$ coupling mobility) does not depend on membrane bending properties.     
In the vanishing frequency limit, or  equivalently for infinite membrane shear and bending moduli, we recover the coupling pair-mobility functions near a hard wall, with no-slip boundary conditions.
Specifically,
\begin{subequations}\label{DeltaMu_TR_Pair_hard} 
 \begin{align}
\lim_{\beta,\betaB\to 0}\frac{\mu_{xy}^{tr,  \mathrm{P}}}{\mu_0^{tr}} &= -\frac{9}{4} \frac{\xi^{1/2}}{(1+\xi)^{5/2}} \, \sigma^2
- \frac{3}{2} \frac{\xi^{1/2} (\xi-4)}{(1+\xi)^{7/2}} \, \sigma^4 \, , \label{DeltaMu_TR_XY_Pair_hard} \\
\lim_{\beta,\betaB\to 0} \frac{\mu_{yx}^{tr,  \mathrm{P}}}{\mu_0^{tr}} &= \frac{3}{2} \frac{\xi^{1/2}}{(1+\xi)^{5/2}} \, \sigma^4 \, , \label{DeltaMu_TR_YX_Pair_hard} \\
\lim_{\beta,\betaB\to 0}\frac{\mu_{yz}^{tr,  \mathrm{P}}}{\mu_0^{tr}} &= \frac{3}{4} \frac{\sigma^2}{(1+\xi)^{3/2}} \, , \label{DeltaMu_TR_YZ_Pair_hard} \\
\lim_{\beta,\betaB\to 0}\frac{\mu_{zy}^{tr,  \mathrm{P}}}{\mu_0^{tr}} &= -\frac{3}{4} \frac{1+4\xi}{(1+\xi)^{5/2}} \, \sigma^2
+\frac{3}{2} \frac{4\xi-1}{(1+\xi)^{7/2}} \, \sigma^4 \, , \label{DeltaMu_TR_ZY_Pair_hard} 
\end{align}
\end{subequations}
in agreement with the results by Swan and Brady~\cite{swan07}.
Note that the components $xy$ and $zy$ keep a negative sign and that $xy$ and $yz$ keep a positive sign in the physical range of parameters in which $\epsilon \in [0,1]$ and $\sigma \in [0, \tfrac{1}{2}]$.

By considering independently the shear and bending contributions to the pair-mobility corrections from Eqs.~\eqref{DeltaMu_TR_Pair}  and taking the limit of vanishing frequency, we obtain for the $xy$ component
\begin{subequations}
 \begin{align}
 \lim_{\beta\to 0} \frac{\mu_{xy, \mathrm{S}}^{tr,  \mathrm{P}}}{\mu_0^{tr}} &= -\frac{3}{8}\frac{\xi^{1/2}(\xi+4)}{(1+\xi)^{5/2}} \, \sigma^2 -\frac{3}{4} \frac{\xi^{1/2}(\xi-4)}{(1+\xi)^{7/2}} \, \sigma^4 \, ,  \\
 \lim_{\betaB\to 0} \frac{\mu_{xy, \mathrm{B}}^{tr,  \mathrm{P}}}{\mu_0^{tr}} &= \frac{3}{8}\frac{\xi^{1/2}(\xi-2)}{(1+\xi)^{5/2}} \, \sigma^2-\frac{3}{4} \frac{\xi^{1/2}(\xi-4)}{(1+\xi)^{7/2}} \, \sigma^4 \, , \label{DeltaMu_TR_XY_Pair_hard_Ben}
\end{align}
\end{subequations}
leading to Eq.~\eqref{DeltaMu_TR_XY_Pair_hard} after summing up both contributions.
It can be shown that the shear-related part is negative whereas the bending related part undergoes a change of sign.
By equating Eq.~\eqref{DeltaMu_TR_XY_Pair_hard_Ben} to zero and solving perturbatively for $\epsilon$, the threshold line where the bending  contribution changes sign is given in a power series of $\sigma$ by
\begin{equation}
\epsilon_\mathrm{th} = \sqrt{2} \sigma \left( 1+\frac{\sigma^2}{3} + \frac{29}{54} \, \sigma^4 \right) + \bigO (\sigma^7) \, .
\end{equation}
Hence, for $\epsilon > \epsilon_\mathrm{th}$, the bending-related part in the coupling mobility is negative whereas it is positive for $\epsilon < \epsilon_\mathrm{th}$.

Next, considering the shear and bending contributions to the component $yx$, we obtain
\begin{subequations}
 \begin{align}
 \lim_{\beta\to 0} \frac{\mu_{yx, \mathrm{S}}^{tr,  \mathrm{P}}}{\mu_0^{tr}} &= \frac{3}{8}\frac{\xi^{1/2}}{(1+\xi)^{3/2}} \, \sigma^2 + \frac{3}{4} \frac{\xi^{1/2}}{(1+\xi)^{5/2}} \, \sigma^4 \, , \\
 \lim_{\betaB\to 0} \frac{\mu_{yx, \mathrm{B}}^{tr,  \mathrm{P}}}{\mu_0^{tr}} &=  -\frac{3}{8}\frac{\xi^{1/2}}{(1+\xi)^{3/2}} \, \sigma^2 + \frac{3}{4} \frac{\xi^{1/2}}{(1+\xi)^{5/2}} \, \sigma^4 \, , 
\end{align}
\end{subequations}
which keep positive and negative signs, respectively, leading to Eq.~\eqref{DeltaMu_TR_YX_Pair_hard} by adding both contributions.
Finally, for the $yz$ component we get
\begin{subequations}
 \begin{align}
 \lim_{\beta\to 0} \frac{\mu_{zy, \mathrm{S}}^{tr,  \mathrm{P}}}{\mu_0^{tr}} &= -\frac{9}{8}\frac{\xi}{(1+\xi)^{5/2}}\, \sigma^2 +\frac{3}{4}\frac{4\xi-1}{(1+\xi)^{7/2}}\, \sigma^4  \\
 \lim_{\betaB\to 0} \frac{\mu_{zy, \mathrm{B}}^{tr,  \mathrm{P}}}{\mu_0^{tr}} &= -\frac{3}{8}\frac{2+5\xi}{(1+\xi)^{5/2}}\, \sigma^2 +\frac{3}{4} \frac{4\xi-1}{(1+\xi)^{7/2}} \, \sigma^4 \, , 
\end{align}
\end{subequations}
both of which are negative valued, leading together to Eq.~\eqref{DeltaMu_TR_ZY_Pair_hard}.

Fig.~\ref{deltaMu_TR_Pair} shows the $tr$ and $rt$ coupling pair mobilities versus the scaled frequency for a pair of particles located above the elastic membrane at $z_0 = 2a$, far apart from each other at a distance $h=4a$.
Membrane shear manifests itself in a more pronounced way for the components $xy$, $yx$, and $yz$, whereas the effect of bending is more significant for the $zy$ component. 
The simulation results are consistent with the fact that the $tr$ and $rt$ coupling mobility tensors are the transpose of each other, as required by the overall symmetry of the mobility matrix.
A very good agreement is obtained between theoretical predictions and boundary integral (BIM) simulations.

%\subsubsection{Rotational mobilities}

We now turn to the rotational pair mobility near an elastic membrane.
In a bulk fluid, the particle rotational mobilities are obtained by inserting the infinite-space Green's function (Oseen tensor) given by Eq.~\eqref{infiniteSpaceGreensFunction} into Eq.~\eqref{particlePairMobility_rot} as
\begin{equation}
\frac{\mu_{xx}^{rr, \mathrm{P}}}{\mu_0^{rr}} = \sigma^3 \, , \qquad 
\frac{\mu_{yy}^{rr, \mathrm{P}}}{\mu_0^{rr}} = \frac{\mu_{zz}^{rr, \mathrm{P}}}{\mu_0^{rr}} = -\frac{1}{2} \, \sigma^3 \, , 
\end{equation}
where $\mu_0^{rr} = 1/(8\pi\eta a^3)$ is the rotational bulk mobility. 
Clearly, the two particles undergo a rotation in the same direction along their line of centers but in opposite direction for the rotation about a line perpendicular to the line of centers, if a torque is exerted on only one of them.
Moreover, the rotational pair mobility along the line of centers connecting the two particles is twice larger in magnitude than the rotational pair mobility for the perpendicular case.

The components of the correction to the rotational mobility are obtained 
%by inserting Eqs.~\eqref{greenFunctions} into 
from Eq.~\eqref{particlePairMobility_rot} as
\begin{subequations}\label{DeltaMu_RR_Pair}
 \begin{align}
\frac{\Delta\mu_{xx}^{rr, \mathrm{P}}}{\mu_0^{rr, \mathrm{P}}} &= \int_0^\infty
\frac{2\sigma^3 u^2}{\xi} \bigg( \frac{B}{\xi^{1/2}} \frac{\varphi}{Bu+i\beta}  \notag \\
&-4i\chi_1 \left( \frac{1}{2iu-\beta} + \frac{4u^2}{8iu^3 - \betaB^3} \right)
\bigg) e^{-2u} \, \Intd u \, , \label{DeltaMu_RR_XX_Pair} \\
\frac{\Delta\mu_{xz}^{rr, \mathrm{P}}}{\mu_0^{rr}} &= \int_0^\infty
\frac{4 B \sigma^3 u^3 \chi_1}{\xi^{3/2}} \frac{e^{-2u}}{Bu + i\beta}  \, \Intd u \, , \label{DeltaMu_RR_XZ_Pair} \\
\frac{\Delta\mu_{yy}^{rr, \mathrm{P}}}{\mu_0^{rr}} &= \int_0^\infty
\frac{2\sigma^3 u^2}{\xi} \left( 
\frac{4i\varphi}{\xi^{1/2}} \bigg( \frac{1}{2iu-\beta} + \frac{4u^2}{8iu^3-\betaB^3} \right)  \notag \\
&-\frac{B \chi_1}{Bu+i\beta}
\bigg) e^{-2u} \, \Intd u \, , \label{DeltaMu_RR_YY_Pair} \\
\frac{\Delta\mu_{zz}^{rr, \mathrm{P}}}{\mu_0^{rr}} &= \int_0^\infty
-\frac{4B\sigma^3 u^3 \chi_0}{\xi^{3/2}} \frac{e^{-2u}}{Bu+i\beta}  \, \Intd u \, . \label{DeltaMu_RR_ZZ_Pair}
\end{align}
\end{subequations}
Similarly, the terms involving $\beta$ and $\betaB$ are related to shear/area dilatation and bending respectively.
It can remarkably be seen that the components $xz$ and $zz$ depend on membrane shear only.
In particular, the correction near a no-slip hard wall is recovered in the zero frequency limit to obtain
\begin{subequations}\label{DeltaMu_RR_Pair_hard}
 \begin{align}
\frac{\Delta\mu_{xx}^{rr, \mathrm{P}}}{\mu_0^{rr}} &= -\frac{1}{2} \frac{2+5\xi}{(1+\xi)^{5/2}} \, \sigma^3 \, , \label{DeltaMu_RR_XX_Pair_hard} \\
\frac{\Delta\mu_{xz}^{rr, \mathrm{P}}}{\mu_0^{rr}} &= \frac{3}{2} \frac{\xi^{1/2}}{(1+\xi)^{5/2}} \, \sigma^3 \, , \label{DeltaMu_RR_XZ_Pair_hard} \\
\frac{\Delta\mu_{yy}^{rr,  \mathrm{P}}}{\mu_0^{rr}} &= -\frac{1}{2} \frac{5\xi-7}{(1+\xi)^{5/2}} \, \sigma^3 \, , \label{DeltaMu_RR_YY_Pair_hard} \\
\frac{\Delta\mu_{zz}^{rr,  \mathrm{P}}}{\mu_0^{rr}} &= -\frac{1}{2} \frac{2\xi-1}{(1+\xi)^{5/2}} \, \sigma^3 \, , \label{DeltaMu_RR_ZZ_Pair_hard}
\end{align}
\end{subequations}
in agreement with the results by Swan and Brady~\cite{swan07}.
Interestingly, the components $yy$ and $zz$ undergo a change of sign for $\xi=7/5$ and $\xi=1/2$, respectively.  
By considering the shear and bending contributions to the pair-mobility corrections independently, from Eqs.~\eqref{DeltaMu_RR_Pair}, and taking the vanishing frequency limit, we obtain for the $xx$ component
\begin{subequations}
 \begin{align}
 \lim_{\beta\to 0} \frac{\mu_{xx, \mathrm{S}}^{rr,  \mathrm{P}}}{\mu_0^{tr}} &=  -\frac{3}{2} \frac{\xi}{(1+\xi)^{5/2}} \, \sigma^3 \, , \\
 \lim_{\betaB\to 0} \frac{\mu_{xx, \mathrm{B}}^{rr,  \mathrm{P}}}{\mu_0^{tr}} &= -\frac{\sigma^3}{(1+\xi)^{3/2}} \, , 
\end{align}
\end{subequations}
leading to Eq.~\eqref{DeltaMu_RR_XX_Pair_hard} after summing up both contributions.
For the component $yy$ we obtain
\begin{subequations}
 \begin{align}
 \lim_{\beta\to 0} \frac{\mu_{yy, \mathrm{S}}^{rr,  \mathrm{P}}}{\mu_0^{tr}} &=  -\frac{3}{2} \frac{\xi-1}{(1+\xi)^{5/2}} \, \sigma^3 \, , \\
 \lim_{\betaB\to 0} \frac{\mu_{yy, \mathrm{B}}^{rr,  \mathrm{P}}}{\mu_0^{tr}} &= -\frac{\xi-2}{(1+\xi)^{5/2}} \, \sigma^3 \, , 
\end{align}
\end{subequations}
leading to Eq.~\eqref{DeltaMu_RR_YY_Pair_hard}.
Accordingly, the shear and bending related parts in the steady limit vanish for $\xi=1$ and $\xi=2$, respectively.

In Fig.~\ref{deltaMu_RR_Pair}, we show the particle scaled rotational pair mobility functions versus the scaled frequency using the same parameters as in Fig.~\ref{deltaMu_TR_Pair}, i.e.\ for a distance from the membranes $z_0 = 2a$ and an interparticle distance $h = 4a$.
As already mentioned, the components $xz$ and $zz$ depend solely on membrane shear resistance whereas both shear and bending manifest themselves for $xx$ and $yy$ components.
As $\xi=1$, the shear-related part in the $yy$ mobility vanishes in the zero frequency limit, and the behavior in the low frequency regime is mainly bending-dominated.
Since the rotational pair mobilities exhibit a scaling as $\sigma^3$, we observe that the corrections are significantly small as compared to the coupling pair mobilities.

\subsection{Doublet of two counterrotating spheres}

\begin{figure}
\begin{center}
\includegraphics[width=0.45\textwidth]{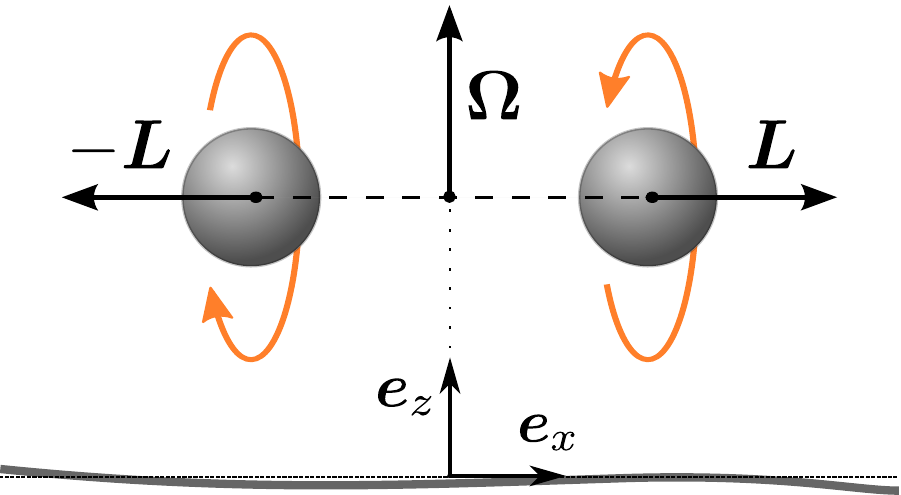}
\caption{Illustration of the two spheres forming a torque-free dimer. Torques $\pm\vect{L}$ of the same magnitude and opposite directions along the center-to-center line are applied to the spheres. 
As a result, the membrane-induced rotation--translation coupling leads to a collective rotation about the axis normal to the membrane with an angular velocity~$\boldsymbol{\Omega}$.
As the $zx$ components of the coupling mobilities vanish, the doublet remains parallel at the same height above the undistorted membrane.} 
\label{corotatingPair}
\end{center}
\end{figure}

%{In order to understand the torque- and force-free behavior of any composite system, we need to employ the hydrodynamic mobility matrix. This is determined by considering the forced motion of the particle, i.e. its response to an external force and torque. These might be of arbitrary origin. In a typical experimental situation of motion of a microparticle close to an elastic interface this could be gravity or optical tweezers. In the theoretical model the effect of this force is induced motion (translational and rotational) which we compute in our considerations.}

To elucidate the effect and role of the change of sign observed in the particle self  and pair mobilities, we consider an example involving the co-rotation within a doublet of particles close to an elastic membrane (see Fig. \ref{corotatingPair}).
We impose external torques, equal in magnitude but oppositely oriented on the pair of particles along the line of centers, causing the particles to rotate in opposite directions.
This set-up may serve as a simple model system to study in an isolated way the rotational effects arising, e.g., during the self-propelled motion of certain types of bacterial microswimmers near an elastic membrane. For example, the bacterium \textit{E.\ coli} propels by rotating a bundle of helicoidal flagella anchored to the cell body. This rotational motion leads to a counterrotation of its actual cell body in the suspending fluid, guaranteeing an overall torque-free motion of the whole microswimmer. Near fluid-solid or fluid-fluid substrates, such rotational properties can qualitatively affect the overall bacterial motion, leading, e.g., to circular trajectories \cite{lauga06, di-leonardo11, daddi18}. In corresponding theoretical studies of bacterial motion, the involved counterrotations have previously been included by an overall torque-free doublet of two antiparallel torques of equal magnitude \cite{lopez14}, similarly to the discretization by our two counterrotating particles in Fig.~\ref{corotatingPair}.

Due to the hydrodynamic coupling between the two counterrotating particles and the membrane, the two particles undergo circular motion along the direction perpendicular to the line of centers.
Accordingly, an induced rotational motion occurs about the center of mass of the doublet with an angular velocity~$\vect{\Omega}$ along the $z$ direction and a rotation rate 
\begin{equation}
{\Omega} = -\frac{2L}{h} \left( \mu_{yx}^{tr, \mathrm{S}} - \mu_{yx}^{tr, \mathrm{P}} \right) \, , \label{omega_def}
\end{equation}
where for the calculation the external torques are applied on both particles along the $x$ axis such that ${L_\lambda}_x = -{L_\gamma}_x = L (t)$.
In fact, Eq.~\eqref{omega_def} holds in the small deformation regime
in which the membrane remains almost planar. Since the steady
state of the membrane deformation is reached quickly, i.e., the memory of the membrane decays significantly quicker than the doublet rotates, we may for our calculation consider the doublet as
oriented along the $x$ axis during the whole time scale relevant for our analytical description.

In the frequency domain, the rotation rate can conveniently be cast in the following generic form 
\begin{equation}
{\Omega}(\omega) = L(\omega) \int_0^\infty \frac{\varphi_1(u)}{\varphi_2(u) + i\omega T} \, \Intd u \, , 
\end{equation}
where the integral represents either the shear- or bending-related parts.
Here $\varphi_2(u) \in \{ u, 2u/B\}$ for the shear contribution and $\varphi_2(u)=u^3$ for bending.
Moreover, $\varphi_1(u)$ is a real function that does not depend on frequency.
We consider now a Heaviside-type function for the torque for which $L(t) = L_0 \, \theta(t)$, the temporal Fourier transform of which to the frequency domain reads $L(\omega) = \left( \pi \delta(\omega)-i/\omega \right) L_0$, with $\delta(\omega)$ the Dirac delta function.
Then the time-dependent rotation rate reads
\begin{equation}
\frac{\Omega(t)}{L_0} = \theta(t) \int_0^\infty \frac{\varphi_1(u)}{\varphi_2(u)} \left( 1-e^{-\varphi_2 (u) \tau} \right) \, \Intd u \, , \label{timeDependentRotationRate}
\end{equation}
wherein $\tau := t/T$ is a scaled time. 
In the steady limit, for which $\tau \to \infty$, the rotation rate can be written in a scaled form as
\begin{equation}
\lim_{\tau\to \infty} \frac{\Omega}{\mu_0^{rr} L_0} = \sigma \epsilon^4 \left(  \frac{8\sigma^5}{\rho^5} -\frac{1}{4} \right) \, , \label{rotationRate_Both}
\end{equation}
where $\rho^2 = \epsilon^2+4\sigma^2$.  
Now, by considering an idealized membrane with pure shear or pure bending rigidities, we obtain
\begin{subequations}
 \begin{align}
\lim_{\tau\to \infty} \frac{\Omega_\mathrm{S}}{\mu_0^{rr} L_0} &= \sigma \epsilon^2 \left( -\frac{1}{4}-\frac{\epsilon^2}{8}+\frac{2\sigma^3}{\rho^3} \left( 1+\frac{2\sigma^2 \epsilon^2}{\rho^2} \right) \right) \, , \label{rotationRate_Shearing} \\
\lim_{\tau\to \infty} \frac{\Omega_\mathrm{B}}{\mu_0^{rr} L_0} &= \sigma \epsilon^2 \left( \frac{1}{4}-\frac{\epsilon^2}{8}-\frac{2\sigma^3}{\rho^3} \left( 1-\frac{2\sigma^2 \epsilon^2}{\rho^2} \right) \right) \, , \label{rotationRate_Bending}
\end{align}
\end{subequations}
leading to Eq.~\eqref{rotationRate_Both} after summing up both contributions. The steady rotation of a torque-free doublet about its center near a membrane with pure shear is of the same sense as near a hard wall.
The rotation is, however, found to be of opposite sense near a membrane with pure bending rigidity.
We note that since the $zx$ components of the $tr$-coupling self and pair mobilities vanish, the pair remain at the same height during its rotational motion above the membrane.

\begin{figure}
\begin{center}
% \scalebox{0.95}{\input{Pics/steadyMotion}}
\includegraphics[scale=0.95]{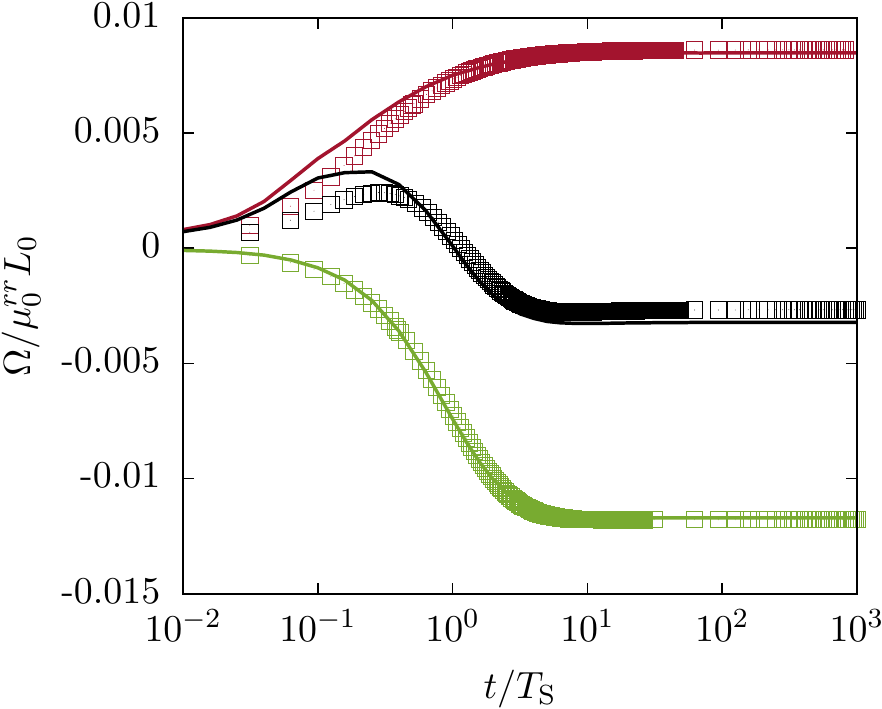}
\caption{(Color online) Scaled rotation rate of the doublet in Fig.~\ref{corotatingPair} versus the scaled time near a membrane of pure shear (green), pure bending (red) and both rigidities (black). Analytical predictions correspond to Eq.~\eqref{timeDependentRotationRate} and symbols refer to boundary integral simulations. Here we use the same parameters as in Fig.~\ref{deltaMu_TR_Pair} for $z_0 = 2a$ and $h=4a$. The time scale of the induced rotation of the pair is considerably slower than that related to the internal rotation of each sphere under the applied torque. } 
\label{steadyMotion}
\end{center}
\end{figure}

In Fig.~\ref{steadyMotion} we present the time-dependent rotation rate of the doublet rotating under a constant external pair of torques exerted along the line of centers, near a membrane with shear-only (green), bending-only (red), or both rigidities (black), as predicted theoretically by Eq.~\eqref{timeDependentRotationRate}.
We observe that at smaller time scales, for which $t/T_\mathrm{S} \ll 1$, the rotation rates amount to small values because the doublet does not yet perceive the presence of the membrane at these short time scales.
As the time progresses, the rotation rates asymptotically approach the values predicted in the steady limit. The resulting rotational motion is slow compared to the angular velocity of each of the spheres under the applied torque. This is due to the weak nature of translational-rotational coupling of a sphere close to a surface, as seen in theoretical calculations \cite{cichocki98} and numerical simulations for a no-slip wall \cite{swan07}.
For a membrane with both shear and bending rigidities, the effect of shear is noticeably more pronounced, leading to the same sense of rotation as predicted near a no-slip wall.
However, for a membrane with pure bending rigidity, such as that of a fluid vesicle~\cite{abreu14}, the steady-sate rotation rate is of positive sign and therefore the pair undergoes rotation of the opposite sense. 
This interesting behavior can alter the near membrane dynamics and behavior of force- and torque-free flagellated bacteria and swimming microorganisms that use helical propulsion as a locomotion strategy~\cite{lauga06, lopez14}.

%{The torque- and force-free conditions come about only to illustrate the motion of a co-rotating doublet of spheres but the results discussed above can be applied to a wider class of forced systems in which we trace the effect of the presence of the membrane on the motion of a spherical particle due to the motion of its neighbor.}

\section{Conclusions}\label{conclusions}

In this paper, we have studied analytically the translation--rotation coupling and rotational hydrodynamic mobilities of a pair of particles moving close to {an elastic membrane} that exhibits resistance towards shear and bending.
We have modeled the elastic membrane by combining the Skalak model for the in-plane shear resistance and the Helfrich model for the out-of-plane bending resistance.
{For example, membranes of red blood cells can be described accordingly.}
For a vanishing actuation frequency or equivalently for higher membrane shear and bending moduli, our results perfectly coincide with those predicted near a hard wall with no-slip boundary conditions.

Using the multipole expansion and \Faxen's theorems, we have expressed the leading order coupling and rotational self- and pair-mobility functions as power series of the ratio between particle radius and membrane distance as well as between radius and interparticle distance.
We have found that the shear- and bending-related contributions may manifest themselves in a supportive or suppressive manner, depending on the membrane properties and the geometric configuration of the particle-membrane system.
As a model system to study the rotational effects involved in certain types of bacterial locomotion, we have studied the rotational dynamics of a torque-free doublet of two counterrotating particles in close vicinity to an elastic membrane. We find that the magnitude and direction of rotation under parallel alignment with the undistorted membrane in the steady limit strongly depend on membrane properties: 
A shear-only membrane rigidity leads to a rotation of the same sense as near a hard wall, opposite to the one near a bending-only membrane rigidity.
Finally, we have verified our theoretical predictions via numerical simulations using a completed double boundary integral method.
A very good agreement is observed in this way.
Our analytically-computed mobility functions may find applications, for instance, as a basis for Brownian simulation studies of colloidal suspensions near planar elastic confinements.

In view of experimental developments involving controlled manipulation of particles in a fluid using optical trapping techniques \cite{reichert04}, an experimental verification of the results presented in the paper might be possible. It would be particularly interesting to explore the dynamics of a rotation-based microswimmer close to the membrane. We have quantified the rotational motion induced by the presence  of the membrane and related it to its elastic properties. These predictions could be tested in a system involving a robotic swimmer in a viscous fluid, perhaps also in larger-scale experiments, as long as the low-Reynolds-number conditions are satisfied.

\section*{Acknowledgements}
The authors gratefully acknowledge support from the DFG (Deutsche Forschungsgemeinschaft) within the projects DA~2107/1-1, ME~3571/2-2, and LO~418/17-2.
SG and ADMI thank the Volkswagen Foundation for financial support and acknowledge the Gauss Center for Supercomputing e.V. for providing computing time on the GCS Supercomputer SuperMUC at Leibniz Supercomputing Center. 
The work has been funded in part by the Ministry of Science and Higher Education of Poland via a Mobility Plus Fellowship  (ML),  and the Foundation for Polish Science within the START programme (ML).
We acknowledge partial support from the COST Action MP1305, supported by COST (European Cooperation in Science and Technology).

\appendix

\section{Green's tensor for a membrane-bounded fluid}\label{appA}
The Green's functions can conveniently be computed using a two-dimensional Fourier transform technique~\cite{bickel07, daddi18jpcm} and appropriately applying the no-slip boundary conditions stemming from shear and bending deformations of the membrane.  

For a point-force exerted at position~$\X_\lambda$ above the membrane, the Green's functions can be expressed in terms of infinite integrals over the wavenumber $q$ as
\begin{subequations}\label{greenFunctions}
\begin{align}
\G_{xx} (\vect{r}, \vect{r}_\lambda, \omega) &= \frac{1}{4\pi}  \int_0^{\infty} 
\bigg( \tilde{\mathcal{G}}_{+} (q,z,z_0,\omega) J_0 (\rho_\lambda q)  \notag \\
  &+   \tilde{\mathcal{G}}_{-} (q,z,z_0,\omega) J_2 (\rho_\lambda q) \cos 2\theta_\lambda \bigg) q \, \Intd q \, ,   \notag \\
\G_{yy} (\vect{r}, \vect{r}_\lambda,\omega) &= \frac{1}{4\pi}  \int_0^{\infty} \bigg( \tilde{\mathcal{G}}_{+} (q,z,z_0,\omega) J_0 (\rho_\lambda q)  \notag \\
  &-   \tilde{\mathcal{G}}_{-} (q,z,z_0,\omega) J_2 (\rho_\lambda q) \cos 2\theta_\lambda \bigg) q \, \Intd q \, ,   \notag \\
{\G}_{zz} (\vect{r}, \vect{r}_\lambda, \omega) &= \frac{1}{2\pi}
\int_{0}^{\infty}  \tilde{\mathcal{G}}_{zz} (q,z,z_0,\omega) J_0 (\rho_\lambda q) q \, \Intd q \, , \notag \\
\G_{xy} (\vect{r}, \vect{r}_\lambda,\omega) &= \frac{\sin 2\theta_\lambda}{4\pi} \int_0^\infty 
\tilde{\mathcal{G}}_{-} (q,z,z_0,\omega) J_2 (\rho_\lambda q) q \, \Intd q \, , \notag \\
\G_{rz} (\vect{r}, \vect{r}_\lambda,\omega) &= \frac{i}{2\pi} \int_{0}^{\infty} 
\tilde{\mathcal{G}}_{lz} (q,z,z_0,\omega) J_1 (\rho_\lambda q) q \, \Intd q \, , \notag \\
\G_{zr} (\vect{r}, \vect{r}_\lambda,\omega) &= \frac{i}{2\pi} \int_{0}^{\infty} 
\tilde{\mathcal{G}}_{zl} (q,z,z_0,\omega) J_1 (\rho_\lambda q) q \, \Intd q \, , \notag
\end{align}
\end{subequations}
where $\rho_\lambda^2 := {(x-x_\lambda)^2 + y^2}$ and $\theta_\lambda := \arctan (y/(x-x_\lambda))$ being the polar angle.
Here $J_n$ denotes the Bessel function of the first kind of order $n$~\cite{abramowitz72}.
Moreover, 
\begin{subequations}
\begin{align}
\tilde{\mathcal{G}}_{zz}  &= \frac{1}{4 \eta q} 
\bigg(
\left( 1+q|z - z_0| \right) e^{-q|z-z_0|}   \notag \\
&+ \left( \frac{i\alpha z z_0 q^3}{1-i\alpha q} + \frac{i\alpha_\mathrm{B}^3 q^3 (1+qz)(1+q z_0)}{1-i\alpha_\mathrm{B}^3 q^3} \right) e^{-q(z+z_0)}  
\bigg) \, ,
\notag  \\
\tilde{\mathcal{G}}_{lz}   &= \frac{i}{4 \eta q} 
\bigg(
-q (z - z_0) e^{-q|z-z_0|}  \notag \\
&+ \bigg( \frac{i\alpha z_0 q^2 (1-qz)}{1-i\alpha q} 
- \frac{i\alpha_\mathrm{B}^3 z q^4 (1+q z_0)}{1-i \alpha_\mathrm{B}^3 q^3} \bigg) e^{-q(z+z_0)}  
\bigg) \, , \nonumber \\
\tilde{\mathcal{G}}_{zl}  &= \frac{{i}}{4 \eta q} 
\bigg(
-q(z - z_0) e^{-q|z-z_0|}  \notag \\
&+ \bigg(- \frac{i\alpha z q^2 (1-q z_0)}{1-i\alpha q} 
+ \frac{i \alpha_\mathrm{B}^3 q^4 z_0 (1+qz)}{1-i \alpha_\mathrm{B}^3 q^3} \bigg) e^{-q(z+z_0)}  
\bigg) \, . \nonumber 
\end{align}
\end{subequations} 
and 
\begin{equation}
\tilde{\mathcal{G}}_{\pm} (q,z,\omega) := \tilde{\mathcal{G}}_{tt}(q,z,\omega) \pm \tilde{\mathcal{G}}_{ll}(q,z,\omega) \, , \notag
\end{equation}
with
\begin{align}
\tilde{\mathcal{G}}_{ll}  &= \frac{1}{4 \eta q} 
\bigg(
(1-q |z - z_0|) e^{-q|z-z_0|}   \notag \\
&+ \left( \frac{i\alpha q (1-q z_0)(1-qz)}{1-i\alpha q} + \frac{i z z_0 \alpha_\mathrm{B}^3 q^5}{1-i \alpha_\mathrm{B}^3 q^3} \right) e^{-q(z+z_0)}  
\bigg) \, , 
\notag
\\
\tilde{\mathcal{G}}_{tt}  &= \frac{1}{2 \eta q} \left( e^{-q|z-z_0|} + \frac{i B\alpha q}{2-i B \alpha q} e^{-q(z+z_0)}  \right) \, ,
\notag
\end{align}
where $\alpha := \kS/(3 B\eta\omega)$ is a characteristic length scale for shear with $B := 2/(1+C)$ as defined in the main body of the paper, and $\alpha_\mathrm{B}^3 := \kB /(4\eta\omega)$ a characteristic cubic length scale for bending.
Thus, the terms involving $\alpha$ and $\alpha_\mathrm{B}^3$ in the above equations are associated with shear and bending, respectively.
Furthermore, the remaining Cartesian components can readily be determined from the usual transformation relations $\G_{xz}=\G_{rz}\cos\theta_\lambda$, $\G_{yz}=\G_{rz}\sin\theta_\lambda$, $\G_{zx}=\G_{zr}\cos\theta_\lambda$, $\G_{zy}=\G_{zr}\sin\theta_\lambda$ and $\G_{yx} = \G_{xy}$.
In the limit of vanishing frequency, the Green's functions near an elastic membrane reduce to the Blake tensor~\cite{blake71} near a rigid no-slip wall, which corresponds to the limit of an immobile and infinitely stiff membrane.

\section{Boundary integral methods}\label{BIM}

In order to assess the accuracy of the multipole expansion approach employed in this paper, we have compared our analytical predictions with fully resolved computer simulations based on the completed double layer boundary integral equation method (CDLBIEM)~\cite{phan93, phan94, kohr04, zhao11, zhao12}.
The method is known to be particularly suited for the simulation of Stokes flows~\cite{pozrikidis01} when both rigid and deformable boundaries are present.
In this way, the translational and rotational velocities of the particles can be determined provided knowledge of the forces and torques exerted on their surfaces.
Hereafter, we briefly provide some technical details regarding the numerical method.

The integral equations for the particle membrane system are expressed as~\cite{daddi17thesis}
\begin{align}
	v_{\beta}(\vect{x}) &= \mathcal{H}_{\beta}(\vect{x}) \,  ,  \quad \vect{x} \in \Sm \, ,  \notag \\
	\frac{1}{2} \, \phi_{\beta}(\vect{x}) + \sum_{\alpha=1}^{6} \varphi_{\beta}^{(\alpha)}(\vect{x}) \langle \vect{\varphi}^{(\alpha)}, \vect{\phi} \rangle &= \mathcal{H}_{\beta}(\vect{x}) \, , \quad \vect{x} \in \Sp \, , \notag 
\end{align}
where $\Sm$ and $\Sp$ denote the surface of the elastic membrane and the particles, respectively.
$\bv$ is the velocity of points belonging to the membrane surface and $\vect{\phi}$ is the so-called double layer density function on the surface of the particles $\Sp$, related to the translational and rotational velocities via 
\begin{subequations}
\begin{align}
\vect{V} (\vect{x}) &= \sum_{\alpha=1}^{3} \vect{\varphi}^{(\alpha)}(\vect{x}) \langle \vect{\varphi}^{(\alpha)}, \vect{\phi} \rangle \, , \quad \vect{x} \in S_{\mathrm{P}} \, , \nonumber \\
\vect{\Omega} (\vect{x}) \times (\vect{x}-\vect{x}_{\mathrm{c}}) &=  \sum_{\alpha=1}^{3}  \vect{\varphi}^{(\alpha+3)}(\vect{x}) \langle \vect{\varphi}^{(\alpha+3)}, \vect{\phi} \rangle \, , \quad \vect{x} \in S_{\mathrm{P}} \, , \nonumber
\end{align}
\end{subequations}
where $\vect{x}_{\mathrm{c}}$ is the particle center and $\vect{\varphi}^{(\alpha)}$ are known vectorial functions that depend on the position of a particle, its surface area, and the moment-of-inertia tensor~\cite{kim13}. 
The brackets stand for the inner product, which is defined as
\begin{equation}
\langle \vect{\varphi}^{(\alpha)}, \vect{\phi} \rangle := \oint_{\Sp} \vect{\varphi}^{(\alpha)} (\vect{y}) \, \mathbf{\cdot} \, \vect{\phi} (\vect{y}) \, \Intd S (\vect{y}) \, ,  \nonumber
\end{equation}
and the function $\mathcal{H}_\beta$ is defined by
\begin{equation}
\begin{split}
 \mathcal{H}_\beta(\vect{x}) &:= -  (\mathcal{N}_{\mathrm{M}} \Delta \vect{f})_\beta (\vect{x}) -  (\mathcal{K}_\mathrm{P} \vect{\phi})_\beta(\vect{x}) \\
		&\quad\,\, + \mathcal{G}_{\beta\mu}^{(0)}(\vect{x},\vect{x}_{\mathrm{c}}) F_\mu 
   + \mathcal{R}_{\beta\mu}^{(0)}(\vect{x},\vect{x}_{\mathrm{c}}) L_\mu
	  \, . \nonumber
\end{split}
\end{equation}
The single and double layer integrals are given by
\begin{subequations}
\begin{align}
(\mathcal{N}_{\mathrm{M}} \Delta \vect{f})_\beta (\vect{x}) &:= \int_{\Sm} \Delta f_\alpha(\vect{y}) \mathcal{G}_{\alpha\beta}^{(0)}(\vect{y}, \vect{x}) \, \Intd S(\vect{y}) \, , \nonumber \\
(\mathcal{K}_\mathrm{P} \vect{\phi})_\beta(\vect{x})        &:= \oint_{\Sp} \phi_\alpha(\vect{y}) \mathcal{T}_{\alpha\beta\mu}^{(0)}(\vect{y},\vect{x}) n_\mu(\vect{y}) \, \Intd S(\vect{y}) \, , \nonumber 
\end{align}
\end{subequations}
with $\vect{n}$ the outer normal vector on the particle surfaces.
Moreover, $\Delta \vect{f}$ is the traction jump of the fluid stress tensor across the membrane, $\mathcal{T}_{\alpha\beta\mu}^{(0)}$ is the stresslet, and $\mathcal{R}_{\beta\mu}^{(0)}$ is the rotlet~\cite{kim13} in an infinite space.
From the instantaneous deformation of the membrane, the traction jump across the membrane $\Delta \vect{f}$ is readily determined from the membrane constitutive models.
For further details with regard to the numerical computation of the traction jumps, we refer the reader to Refs.~\onlinecite{daddi16b, guckenberger16}.

In our simulations, the planar membrane is a flat quadratic surface with a size of $300 a \times 300 a$ and is meshed with 1740 triangles created using the open-source and freely-available software gmsh~\cite{Geuzaine2009}.
The spherical particle is discretized by 320 triangular elements obtained by consecutive refinement of an icosahedron~\cite{krueger11, gekle16, bacher17, guckenberger18}.

For the numerical determination of the particle mobility functions, a harmonic force $\F_\lambda(t) = \vect{A}_\lambda e^{i\omega_0 t}$ or torque $\vect{L}_\lambda(t) = \vect{B}_\lambda e^{i\omega_0 t}$ is exerted at the surface of the particle $\lambda$.
After a brief transient evolution, the translational and rotational velocities of the particle $\gamma$ evolve as $\vect{V}_\gamma (t) = \vect{C}_\gamma e^{i(\omega_0 t + \delta_\gamma)}$ and $\boldsymbol{\Omega}_\gamma (t) = \vect{D}_\gamma e^{i(\omega_0 t + \varphi_\gamma)}$, respectively, and analogously for the particle $\lambda$.
The amplitudes and phase shifts can accurately be determined by a fitting procedure of the numerically recorded velocities using the trust region method \citep{conn00}.
In this way, the $rt$ components can be computed for a torque-free particle as
\begin{equation}
\mu_{\alpha\beta}^{rt, \lambda\lambda} = \frac{{D_\lambda}_\alpha}{{A_\lambda}_\beta} \, e^{i\varphi_\lambda} \, , \qquad
\mu_{\alpha\beta}^{rt, \gamma\lambda} = \frac{{D_\gamma}_\alpha}{{A_\lambda}_\beta} \, e^{i\varphi_\lambda} \,  . \notag
\end{equation}
For a force-free particle, the components $tr$ and $rr$ are computed from
\begin{equation}
\mu_{\alpha\beta}^{tr, \lambda\lambda} = \frac{{C_\lambda}_\alpha}{{B_\lambda}_\beta} \, e^{i\delta_\lambda}  \, , \quad 
\mu_{\alpha\beta}^{rr, \lambda\lambda} = \frac{{D_\lambda}_\alpha}{{B_\lambda}_\beta} \, e^{i\varphi_\lambda} \notag
\end{equation}
for the self mobilities and 
\begin{equation}
\mu_{\alpha\beta}^{tr, \gamma\lambda} = \frac{{C_\gamma}_\alpha}{{B_\lambda}_\beta} \, e^{i\delta_\gamma}  \, , \quad 
\mu_{\alpha\beta}^{rr, \gamma\lambda} = \frac{{D_\gamma}_\alpha}{{B_\lambda}_\beta} \, e^{i\varphi_\gamma} \notag
\end{equation}
for the pair mobilities.

\input{main.bbl}
%\bibliography{biblio} %your .bib file

\end{document}

%% file: commands.tex
\newcommand{\vect}[1]{\boldsymbol{#1}}

\newcommand{\bv}{\vect{v}}
\newcommand{\BV}{\boldsymbol{V}}
\newcommand{\BF}{\boldsymbol{F}}

\newcommand{\BL}{\boldsymbol{L}}

\newcommand{\mi}{\bm{\mu}}

\newcommand{\zBbar}{\overline{z_{\mathrm{B}}}}

\newcommand{\boldNabla}{\boldsymbol{\nabla}}

\newcommand{\kS}{\kappa_\mathrm{S}}
\newcommand{\kA}{\kappa_\mathrm{A}}
\newcommand{\kB}{\kappa_\mathrm{B}}

\newcommand{\betaB}{\beta_\mathrm{B}}

\newcommand{\G}{\mathcal{G}}

\newcommand{\Intd}{\mathrm{d }}

\newcommand{\F}{\vect{F}}

\newcommand{\X}{\vect{x}}

\newcommand{\E}{\operatorname{E}}
\newcommand{\bigO}{\mathcal{O}}

\newcommand{\Sm}{S_{\mathrm{M}}}
\newcommand{\Sp}{S_{\mathrm{P}}}

\newcommand{\Faxen}{Fax\'{e}n}

\newcommand{\Gmatr}{\boldsymbol{\mathcal{G}}}
\newcommand{\Tmatr}{\boldsymbol{\mathcal{T}}}

\newcommand{\zB}{z_{\mathrm{B}}}

\newcommand{\EB}{E_\mathrm{B}}